\documentclass[12pt,preprint]{aastex}
\usepackage{graphicx}

\slugcomment{Draft}

\newcommand{\iraf}{\textsc{iraf}}

\newcommand{\mscred}{\textsc{mscred}}
\newcommand{\sex}{\textsc{sextractor}}
\newcommand\om{\ifmmode \Omega_{\mathrm{M}}\else $\Omega_{\mathrm{M}}$\fi}
\newcommand\ol{\ifmmode \Omega_{\Lambda}\else $\Omega_{\Lambda}$\fi}

\def\ergscmhz{\ifmmode \mathrm{erg}\,\mathrm{s}^{-1}\,\mathrm{cm}^{-2}\,\mathrm{Hz}^{-1}\else $\mathrm{erg}\,\mathrm{s}^{-1}\,\mathrm{cm}^{-2}\,\mathrm{Hz}^{-1}$\fi}

\shorttitle{The Phoenix Deep Survey}
\shortauthors{M. Sullivan et al.}

\begin{document}


\title{The Phoenix Deep Survey: Optical and near infrared imaging catalogs}


\author{M. Sullivan\altaffilmark{1,2}, A. M. Hopkins\altaffilmark{3,9}, 
        J. Afonso\altaffilmark{4}, A. Georgakakis\altaffilmark{5},
        B. Chan\altaffilmark{6},
        L. E. Cram\altaffilmark{7}, B. Mobasher\altaffilmark{8}, C. Almeida\altaffilmark{4}}

\affil{
\begin{enumerate}
\item Department of Physics, University of Durham, South Road, Durham,
 DH1 3LE, UK
\item Department of Astronomy and Astrophysics, University of Toronto, 60 St. George Street, Toronto, ON M5S 3H8, Canada
\item Department of Physics and Astronomy, University of Pittsburgh,
 3941 O'Hara St, Pittsburgh, PA 15206, USA
\item CAAUL, Observat\'orio Astron\'omico de Lisboa, Tapada da
Ajuda, 1349-018 Lisboa, Portugal
\item Institute of Astronomy and Astrophysics, National Observatory of Athens,
 I. Metaxa \& B. Pavlou, Penteli, 15236, Athens, Greece
\item School of Physics, University of Sydney, NSW 2006, Australia
\item Australian Research Council, GPO Box 2702, Canberra ACT 2601, Australia
\item Space Telescope Science Institute, 3700 San Martin Drive,
 Baltimore, MD 21218, USA
\item Hubble Fellow
\end{enumerate}
}
\email{sullivan@astro.utoronto.ca}

\begin{abstract}
  The Phoenix Deep Survey is a multi-wavelength galaxy survey based on
  deep 1.4\,GHz radio imaging (Hopkins et al., 2003). The primary goal
  of this survey is to investigate the properties of star formation in
  galaxies and to trace the evolution in those properties to a
  redshift $z=1$, covering a significant fraction of the age of the
  Universe. By compiling a sample of star-forming galaxies based on
  selection at radio wavelengths we eliminate possible biases due to
  dust obscuration, a significant issue when selecting objects at
  optical and ultraviolet wavelengths. In this paper, we present the
  catalogs and results of deep optical ($UBVRI$) and near-infrared
  ($Ks$) imaging of the deepest region of the existing decimetric radio
  imaging. The observations and data-processing are summarised and the
  construction of the optical source catalogs described, together with
  the details of the identification of candidate optical counterparts
  to the radio catalogs.  Based on our $UBVRIKs$ imaging, photometric
  redshift estimates for the optical counterparts to the radio
  detections are explored.
\end{abstract}


\section{Introduction}
\label{sec:introduction}

The study of galaxy evolution in recent years has included a strong
focus on the star formation properties of galaxies
\citep{1996ApJ...460L...1L,1996MNRAS.283.1388M,1997ApJ...486L..11C,2000ApJ...544..218A,2000ApJ...544..641H,2000MNRAS.312..442S,2000AJ....120.2843H,2002A&A...383..801B,2002AJ....124..675C,2002MNRAS.337..369T,2003ApJ...584..210G}.
Many of these studies draw their galaxy samples primarily from
selection at ultraviolet (UV) or optical wavelengths (for example
broad-band imaging or through the H$\alpha$ emission line), and are
known to be affected by obscuration due to dust intrinsic to the
galaxies of interest. Correcting for the effects of this dust has been
shown to be a complex problem, especially for distant objects
\citep[e.g.][]{2001PASP..113.1449C}; furthermore selection at optical
and UV wavelengths results in samples of star forming systems which
may omit a significant fraction of heavily obscured galaxies
\citep{2002ApJ...581..844S}. A related issue is that the most strongly
star-forming systems are on average the most heavily obscured
\citep{2001AJ....122..288H,2001ApJ...558...72S,2002A&A...383..801B,
  2003astroph0306621H,2003astroph0307175A}, with recent results
suggesting an average Balmer-decrement derived obscuration of
radio-detected systems of $A_{\mathrm{H}\alpha}\sim1.6$
\citep{2003astroph0307175A,2003astroph0306621H}, higher than in UV or
H$\alpha$ selected samples
\citep[$A_{\mathrm{H}\alpha}\simeq1$;][]{1998ApJ...495..691T,2000MNRAS.312..442S}.

Selecting galaxies at radio or far-infrared (IR) wavelengths could
alleviate many of the potential problems associated with selecting
galaxies at optical/UV wavelengths. Far-IR emission is a by-product of
massive star-formation, a result of UV radiation being absorbed by
dust and re-emitted in the 10-300$\mu$m wavelength range. The
existence of a tight and linear relationship between this far-IR
emission and radio luminosity in local galaxies
\citep[e.g.,][]{1985A&A...147L...6D,1991ApJ...376...95C,2003ApJ...586..794B}
suggests a strong link between radio emission and unobscured massive
star-formation, though the specific details of the physical model of
the radio emission are still poorly understood theoretically
\citep{1992ARA&A..30..575C}. Hence, selecting at radio wavelengths can
provide samples of star-forming galaxies without regard to their
intrinsic dust content.

In order to identify a homogeneously selected catalog of star-forming
galaxies, unbiased by obscuration-based selection effects, and
spanning a broad redshift range ($0<z<1$), the Phoenix Deep Survey
(PDS\footnote{see also
  http://www.atnf.csiro.au/people/ahopkins/phoenix/}) relies on a deep
1.4\,GHz mosaic image, based on observations with the Australia
Telescope Compact Array. This has been used to construct one of the
largest existing deep 1.4\,GHz source catalogs
\citep{2003AJ....125..465H} from which a star-forming galaxy
sub-sample can be drawn. The PDS has already been highly successful in
providing a basis for several investigations of the nature of star
forming galaxies and their evolution
\citep[e.g.,][]{1999MNRAS.306..708G,1999MNRAS.310L..15G,1999MNRAS.308...45M,2000ApJS..128..469H,2000A&AS..141...89G,2003astroph0307175A}.
Additionally, 2dF spectroscopic information exists for a large
fraction of the optically brighter optical counterparts
\citep{1999MNRAS.306..708G}.

In this paper, we present the details of deep optical and
near-infrared (NIR) imaging carried out over the deepest portion of
the 1.4\,GHz radio survey. These observations will, in later papers,
be used to investigate the star-formation and dust properties of the
faint radio source population, and the evolution of these properties
out to $z\simeq1$. A plan of this present paper follows. In
Section~\ref{sec:imaging-datasets}, we present the various imaging
datasets and discuss the steps taken to reduce the raw data.
Section~\ref{sec:source-detect-catal} describes the processes involved
in creating robust source catalogs drawn from the optical imaging, and
the technique used to cross-correlate this optical dataset with the
existing radio catalog.  In Section~\ref{sec:photo-z} we investigate
the degree to which photometric redshift estimates can be made for our
radio-selected sample, and in Section~\ref{photometric-properties}, we
discuss the broad photometric properties of the radio-selected sample.
We summarise our results in Section \ref{sec:summary}.  Throughout
this paper we assume a $\ol=0.7$, $\om=0.3$, $h=0.70$ (where
$H_0=100\,h\,\mathrm{km\,s^{-1}\,Mpc}^{-1}$) cosmology.

\section{Imaging datasets}
\label{sec:imaging-datasets}

\begin{figure}
  \centering 
  \plotone{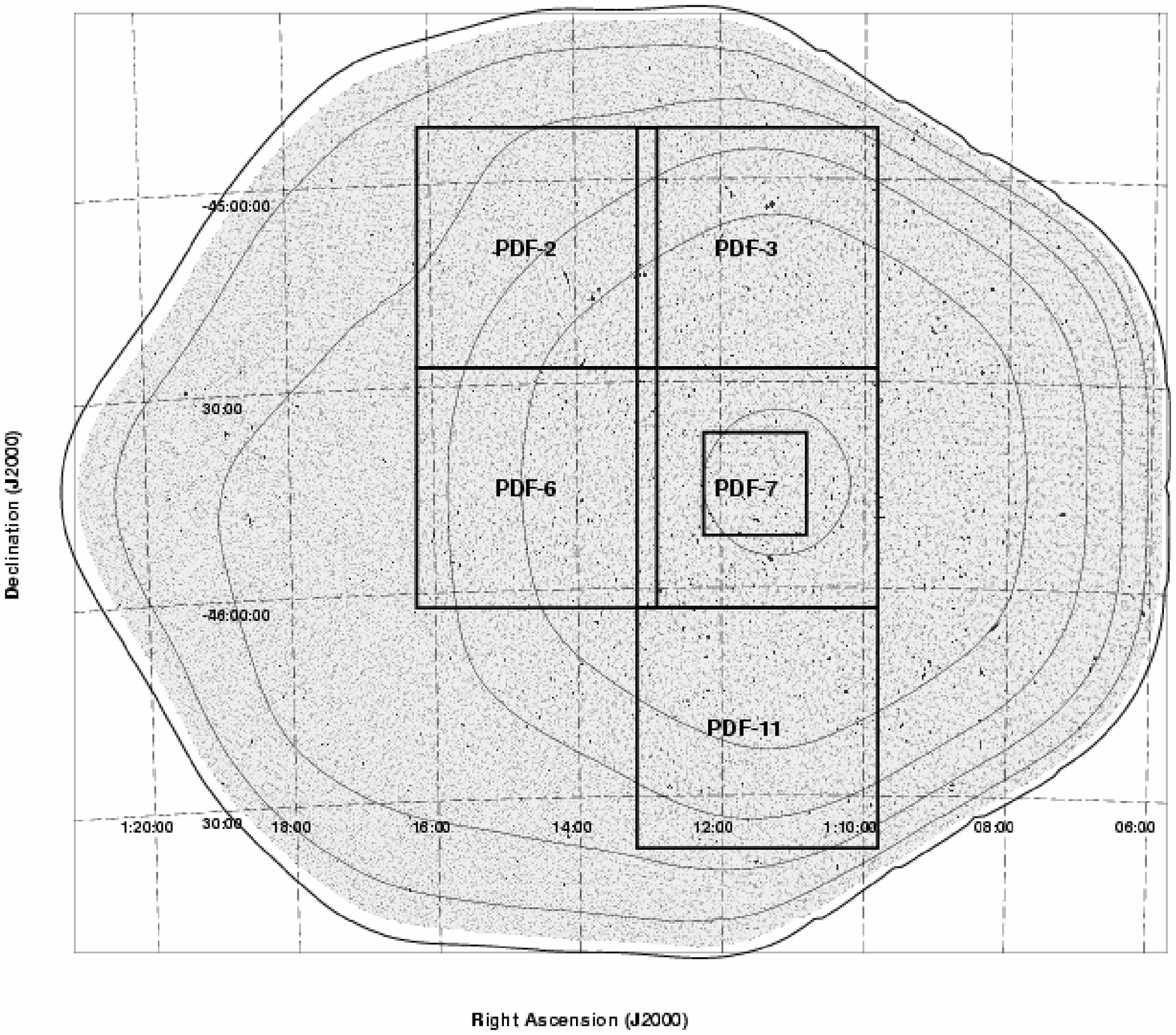}
  \caption{
    The optical mosaicing pattern of the PDS overlaid on the full
    1.4\,GHz ATCA radio map. Contours show the theoretical r.m.s.
    noise level over the radio mosaic, ranging from 10\,$\mu$Jy (inner
    circle) to 0.32\,mJy, in steps of factors of 2. The large
    rectangular boxes show the location of each optical mosaic
    pointing and the associated numbering system. The small rectangle
    denotes the extent of the $Ks$ data.\label{fig:pds_coverage}}
\end{figure}

In this section, we describe the imaging observations of the PDS. This
comprises three primary sets: AAT wide field imager (WFI) optical
$BVRi$ data, CTIO Mosaic-II $U$-band data, and ESO-WFI $U$-band data.
These imaging cameras cover approximately the same area on the sky --
around 0.35\,deg$^2$ -- and hence the same mosaicing pattern was used
at each telescope to facilitate the alignment of data from different
filters.  The extent of the coverage of the radio field and the
numbering of the pointing pattern used in relation to the deep radio
survey can be seen in Figure~\ref{fig:pds_coverage}. In addition to
the optical data, near-IR NTT/SOfI $Ks$ data also exists for part of
the central pointing (see pointing 7 in
Figure~\ref{fig:pds_coverage}), as well as relatively shallow AAT $R$
and $V$-band data ($m_R\simeq22.5$) for most of the PDS field
presented in \citet{1999MNRAS.306..708G}.  Below we describe the
observations and reduction of the new optical datasets; the
observations in this section are summarised in
Table~\ref{tab:optical_observations}.

\subsection{Optical imaging observations}
\label{sec:optical-imaging}

The following observations were taken. Two pointings (7,3) were
observed in $BVRi$ and one (pointing 11) in $BVi$ on the nights of
August 13 and 14 2001, with the WFI camera on the AAT.  The same three
pointings were also observed in $U$ with the Mosaic-II camera on the
CTIO 4m Blanco telescope on September 3rd 2002. Finally, four of the
PDS fields (2,3,6,7) were observed in $U$ with the WFI on the ESO 2.2m
on the night of August 18 2001. All exposure sequences were dithered
to cover the inter-chip spacing; details can be found in
Table~\ref{tab:optical_observations}. All the science observations
were interspersed with observations of standard star fields taken from
\citet{1992AJ....104..340L} -- all the data presented in this paper
were secured during photometric nights.

We will not discuss the reduction of the individual datasets, as these
typically followed the same path; instead we outline the generic
reduction steps below and discuss issues specific to each dataset
reduction where appropriate. All the data were reduced using \iraf\ 
and the National Optical Astronomy Observatories (NOAO) mosaic
reduction software package \mscred\ version 4.7
\citep{1998adass...7...53V}, as well as some custom-written routines
where required. Data were bias-subtracted using overscan regions and
a master bias frame generated on each night of observation. Flat-fielding
was performed using dome-flats ($BVRi$ data) in the
standard manner. For $U$-band data, dome flats provide a poor
flat-field correction and consequently we flat-field these data using
twilight sky flats. Bad-pixel masks are also generated for each image,
containing bad column, saturated stars, and bleed trails identified by
\mscred.

Astrometric solutions were applied to each science frame using the
\iraf\ TNX projection. For CTIO $U$-band data we applied the
distortion correction provided by the observatory. For the AAT and ESO
data we generated our own distortion corrections from the observation
fields and USNO-2 stars. These solutions were typically accurate to
$\simeq0\farcs35$. The astrometrically calibrated frames were then
projected and re-sampled using a sinc function onto a standard linear
world coordinate system (WCS) with a constant pixel-scale, linearly
interpolating over bad-pixels to avoid ringing effects in the
re-sampled data. The re-sampling was performed so that each image for
a particular pointing was placed onto the same pixel coordinate system
-- in general we aligned to the CTIO Mosaic-II system which has the
largest field-of-view and the most precise WCS. The realignment
process is accurate to $0\farcs1$. In order to combine data in each
filter, we determined the sky levels in each image as follows. We ran
\sex\ \citep{1996A&AS..117..393B}, outputting a FITS file object mask.
We then used an iterative technique to determine the modal pixel
value, excluding those pixels lying within object boundaries
determined by the \sex\ mask.  The modal value was subtracted from
each data frame, and the rms dispersion in the sky level recorded.

The individual dither steps at each pointing were then
median-combined, masking cosmic-rays, chip defects and satellite
trails, creating the deepest possible final image for each pointing.
We applied multiplicative scaling factors to each image (calculated
from common USNO-2 stars) to allow for airmass variations between
exposures - zero-point offsets were not required due to the previous
subtraction of the modal sky value. The result is an
astrometrically-calibrated and photometrically-stable combined science
image. Finally, the modal sky value was added back to preserve the
count statistics for later analyses.

The photometric calibration was performed using stars drawn from the
catalog of \citet{1992AJ....104..340L}. These frames were observed at
a variety of airmasses throughout the night, with an effort made to
ensure that standard stars fell on each chip in the mosaic arrays.
Zeropoints were derived from these data in the standard way, and the
large number of stars observed allowed us to fit for both for the
zeropoint and airmass/color terms. We found negligible color terms
(less than 0.02\,mag) in the $BVR$ data relative to the standard
Johnson-Cousins system \citep[e.g.][]{1990PASP..102.1181B}, suggesting
these filter/CCD response systems approximate well to the standard
calibration system. As the $i$-band data, however, were obtained
through an SDSS $i$-band filter and not a Cousins-$I$ filter ($I_c$),
a non-negligible color term was required to align these data into the
$I_c$ system of the standard stars.  We found that to convert our
measured CCD photometry with this filter, $i_{\rm CCD}$, to the
standard Cousins system ($I_c$) required a transformation of the form

\begin{equation}
I_{c}=i_{\rm CCD}-0.194\times(R_{c}-I_{c}).
\end{equation}

\noindent
where $i_{\rm CCD}$ is the calibrated $i$-band magnitude with no color
term applied, and $R_c$ is the calibrated $R_c$-band magnitude.

The $U$-band data also possess small color terms, though this is
harder to calculate due to the limited precision to which
\citet{1992AJ....104..340L} stars have been observed in the $U$.
Indeed, a large uncertainty connected with $U$-band imaging is the
accuracy, and consistency, to which a photometric calibration can be
performed, due to the high sensitivity to even small amounts of
cirrus, and the effect of varying site-to-site extinction laws.

\begin{figure}
  \centering
  \plotone{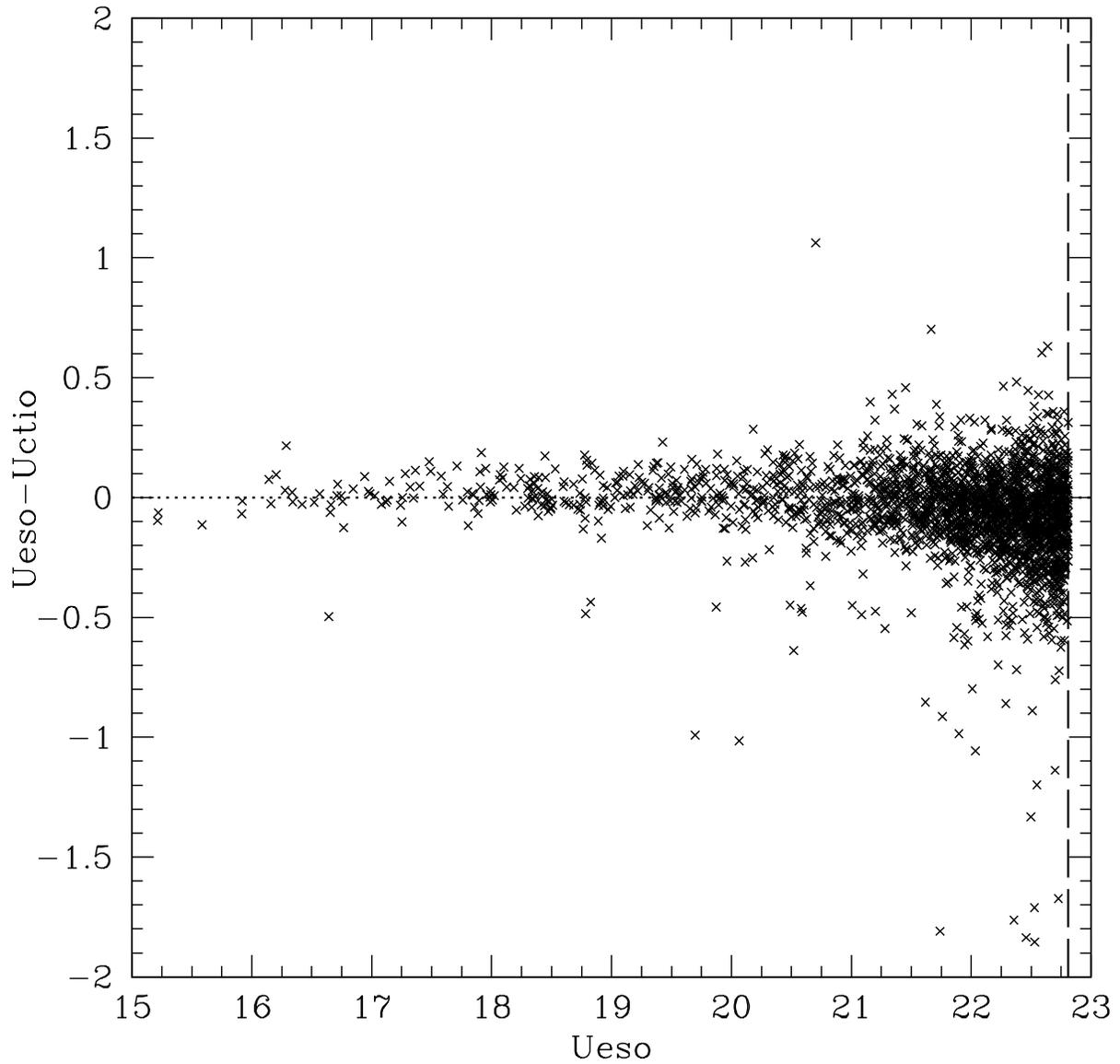}
  \caption{
    A comparison of $3''$ aperture magnitudes for common objects
    between the CTIO and ESO $U$-band data, shown as the difference in
    the magnitudes as a function of the ESO magnitude. The vertical
    dashed line shows the limiting magnitude of the ESO data.The
    figure demonstrates an excellent agreement between $U$-band data
    taken at different sites with different
    telescopes/instruments.\label{fig:uband-ctioeso-compare}}
\end{figure}

We are fortunate, however, to be able to perform an internal test on
the consistency of our $U$-band data. As two of the ESO pointings
overlap with the CTIO $U$-band data (numbers 3 and 7), we can test the
precision of our $U$-band photometric calibration by comparing the
aperture magnitudes of objects common to both datasets. In the
coinciding frames we measure $3''$-diameter aperture magnitudes for
the objects detected by \sex, the comparison of which is shown in
Figure~\ref{fig:uband-ctioeso-compare}. The agreement between the two
datasets is excellent; the median offset is 0.01\,mag for ESO
detections with $U<21$, and -0.03\,mag for all objects which lie above
the limiting magnitude of the ESO data. Given that these data were
taken at different sites with different telescope/instrument
combinations, this represents a highly encouraging consistency check
between our $U$-band datasets. Where common objects are observed
between the two datasets, we always take the CTIO magnitudes as our
default measure due to the greater depth of these data.

\subsection{Near infrared NTT/SOfI observations}
\label{sec:k-band-imaging}

Our NIR imaging data come from the Hawaii HgCdTe 1024$\times$1024
pixel-array SoFI camera on the 3.6m ESO New Technology Telescope
(NTT). The field-of-view was $4.9\arcmin \times 4.9\arcmin$ with a
pixel scale of $0\farcs29$. Nine contiguous pointings, in a $3\times
3$ pattern, were observed over the deepest region of the PDS (a
sub-region of pointing 7; see Figure~\ref{fig:pds_coverage}), during
October 10 and October 11 2000.  At each pointing, 45 dithered 60s
exposures were taken in the $Ks$ filter, except for the central
pointing where this sequence was repeated 4 times. Each 60s exposure
comprised the average of six 10s sub-exposures. All data were taken
under photometric conditions with seeing of less than $1\farcs2$.

The raw data were reduced using \iraf\ following a method developed by
Peter Witchalls and Will Saunders, outlined here. To correct for the
varying pixel responses to light from astronomical objects, an image
of a source with uniform light intensity over the entire field-of-view
is required. Ideally we require two such images. The first is a
standard NIR dome-flat, appropriate for correcting the pixel-to-pixel
variations in response to astronomical objects. This is created by
averaging the subtraction of many ``dome-light-on minus
dome-light-off'' image pairs.  However, this dome-flat is not
appropriate for correcting for the response to variable sky
illumination. It is the highly variable nature of this NIR night-sky that
makes this second flat-field important, as we must flatten the pixel
response to the night-sky before subtracting the sky background.  The
method we use here makes use of a technique which uses the night-sky
itself as this second flat-field.

The procedure is as follows. First, all the raw science frames were
median combined to produce a super-sky image, which was then
subtracted from every raw data frame.  The differences between
consecutive ``supersky-subtracted" science frames was then calculated,
giving the spatial and temporal changes in sky brightness levels from
frame to frame. As these difference frames are noise dominated due to
the small amplitude of the sky fluctuations, they are normalised and
combined to make a high-S/N flat-field. This ``difference-flat" is
then divided into all the science frames, correcting for the variable
pixel response to the sky illumination. 

To subtract the remaining sky level in each individual science frame,
the 8 temporally closest science frames are median-combined (masking
stars and bright objects) and the result subtracted from that science
frame, leaving a sky-subtracted image.  Multiplication by the
difference-flat and then division by the dome-flat then corrects for
the varying pixel-to-pixel response to light from astronomical
objects.

The photometric calibration was performed using infrared standards
from \citet{1998AJ....116.2475P}, observed throughout each night. We
also dithered one standard star across the array as a check on the
flat-fielding quality; the dispersion in the measured magnitudes of
this standard star was $\sim 0.015$\,mag. The photometric calibration
was similar for each night, and we derived a common solution with an
r.m.s. scatter of $\sim 0.02$\,mag.

The sub-images were then registered and combined, firstly for each
individual pointing, and then finally for the full
$14\times14$\,arcmin mosaic, masking chip defects.  Astrometric
calibration was made using the SUPERCOSMOS catalog, with $\sim$40
stars identified and used to build the astrometric solution.  The
final r.m.s. of the solution was $\simeq0\farcs2$, similar to that
found in the optical imaging. As a final step, the $Ks$ mosaic was
re-sampled to the same WCS and pixel scale as the optical data in
pointing 7 for use in constructing the source catalogs.

\subsection{Aligning the magnitude systems}
\label{sec:align-magnitudes}

Each of our datasets have been calibrated to the standard Vega-based
magnitude system using \citet{1992AJ....104..340L} standard star
fields. We prefer, however, to apply further corrections in order to
place our magnitudes onto the easily-interpretable AB-magnitude system
\citep[e.g.][]{1974ApJS...27...21O}, defined as

\begin{equation}
m_{\mathrm{AB}} = -2.5 \log(F) - 48.60
\end{equation}

\noindent
where the flux, $F$, is in units of \ergscmhz. This system allows an
easy conversion into fluxes for any magnitude. Conversions for the
$BVRIKs$ magnitudes were calculated using the standard filter
responses.  For the $U$-band data we use the filter responses as
measured for the system in use at the telescope.  The AB conversions
are listed in Table~\ref{tab:abconversions}, and are similar to
conversions tabulated in the literature \citep{1995PASP..107..945F}.

\section{Source detection and catalog construction}
\label{sec:source-detect-catal}

To make full use of our multi-wavelength dataset, we require robust
and reliable catalogs drawn from each observation set, containing
all objects present in all passbands. In this section, we describe the
technique used to construct the source catalogs from our data.

\subsection{Preliminaries}
\label{sec:preliminaries}

Our principle tool in constructing the optical catalogs is \sex\ 
version 2.2.2 \citep{1996A&AS..117..393B}, which we run from within
custom-written \textsc{perl} routines which handle the various aspects
of the catalog construction. We first determine the seeing in each
data frame using a \textsc{perl} routine, running \sex\ using a high
detection threshold and an estimate of the seeing based on visual
inspection using \textsc{imexamine}. From the catalog of detected,
bright objects, we extract a list of objects classified as stellar
according to the \sex\ neural network classifier. We feed this list to
the \textsc{phot} package in \iraf, fitting Gaussian profiles to each
object, rejecting all objects with fitted ellipticities of $<0.95$ as
well as those which lie near to the edge of a particular image. The
100 objects with the smallest fitted FWHM are extracted, and the
median FWHM value of this list taken as the seeing estimate for that
image. In general, the seeing of all the images is very similar as
conditions were stable during the observations. The exception is the
$Ks$-band data, which was taken in superior seeing conditions. The
measured seeing values are listed in Table~\ref{tab:optical_depths}.
Finally, we convolve each image to the poorest seeing of the
observation set, and generate two catalogs, one containing
seeing-matched data and the other the original data. Again, with the
exception of the $Ks$-band data, this step is not critical.

\subsection{Catalog completeness limits}
\label{sec:catalog-completness}

For the final catalog construction, we also need to estimate the
completeness magnitudes for each image. We estimate this in two ways.
Firstly, we estimate the 5-$\sigma$ limiting magnitude for each image.
We use \sex\ to detect a catalog of objects, sorting the list based on
the errors in the object magnitudes. The 5-$\sigma$ limiting magnitude
is taken as the median magnitude of all objects with magnitude errors
in the range $0.19-0.21$, and the values are listed in
Table~\ref{tab:optical_depths}.

\begin{figure}
  \centering
  \plotone{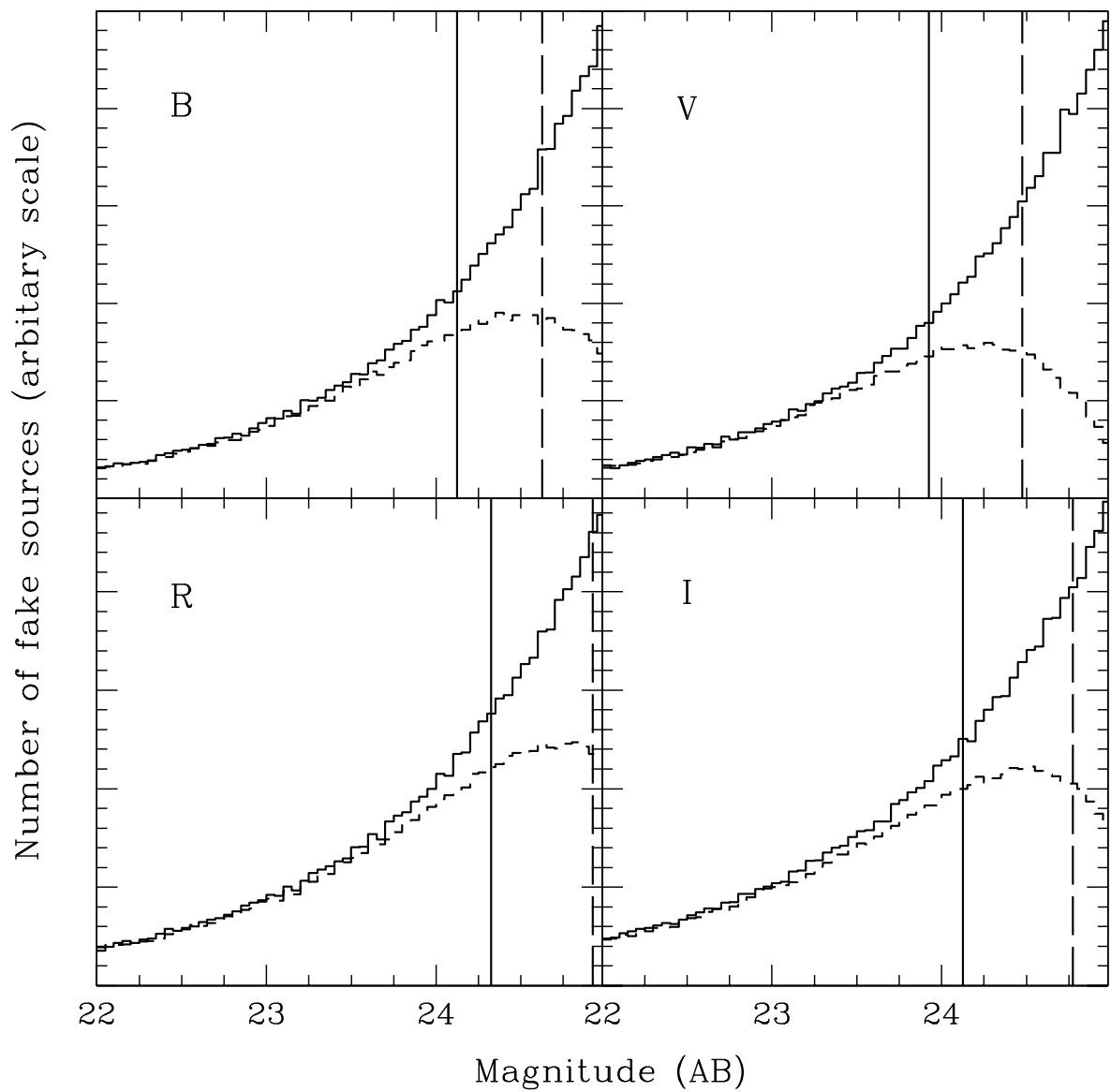}
  \caption{
    Example simulations of the optical completeness limits for the
    central pointing of the PDF survey (pointing PDF-7 in
    Fig.~\ref{fig:pds_coverage}). Solid curves denote the cumulative
    counts for ``fake'' sources placed into the images; the broken
    curve shows the fraction recovered as a function of magnitude. The
    vertical solid and dashed lines show 80\% and 50\% completeness
    magnitudes respectively. See text for details of the simulations.
    \label{fig:completness}}
\end{figure}

The other approach is to estimate 50 and 80\% completeness limits by
inserting fake sources into our imaging data, and measuring the
fraction recovered as a function of magnitude. Our procedure is as
follows. For each filter, we insert into the images sources at random
(but recorded) positions with a magnitude distribution which follows
the galaxy number counts of \citet{2003astro.ph..6254M}, the total
number inserted normalised to the effective area of each pointing.
Each source is inserted with a FWHM determined in
Section~\ref{sec:preliminaries}.  Using \sex, we determine how many of
these sources are recovered as a function of magnitude. This
simulation is repeated 100 times for each pointing/filter combination,
and the magnitude at which 50\% and 80\% of sources are recovered is
recorded (see Figure~\ref{fig:completness} for example histograms of
these distributions from the pointing PDF-7). The estimated
completeness limits for each pointing and filter are listed in
Table~\ref{tab:completeness}.

\subsection{Constructing the detection image}
\label{sec:constr-detect-image}

With large mosaic images across up to six photometric bands,
constructing reliable object catalogs, containing all possible
detections from all available bands, is a complex task. For
example, if we choose to detect objects in only one passband and then
measure magnitudes in the remaining passbands using the positional
parameters derived from the detection image, we will likely (1) exclude
objects with unusual colors, and (2) miss detections that, while not
significant in any one passband, are clear detections when all the
available data is combined. As one of the goals of this project is to
detect the very-faint optical counterparts to a deep radio survey,
having a source catalog with all possible detections is clearly
important. The alternative approach of matching catalogs created for
each individual passband will solve the first problem (though not the
second), but introduces complex uncertainties if different apertures
are used for magnitude measures in different passbands.

We choose to use a modified version of the multicolor object detection
technique of \citet{1999AJ....117...68S}. Essentially, this is a
``$\chi^2$"-technique. We construct a ``detection" image from $N$
different passbands by co-adding, for each pixel, the sky-subtracted
flux ($f-s$) weighted by the dispersion in the sky ($\sigma$)
according to

\begin{equation}
  \label{eq:1}
  \chi^2=\Sigma^{N}_{i=1}(\frac{f_i-s_i}{\sigma_i})^2
\end{equation}

This technique is particularly suited to our dataset, as all the
optical data were taken at the same position in the sky, and hence
the pixels in the $\chi^2$ image have contributions from all the
passbands.

This technique assumes that each image is sky-noise limited, and that
the depth is constant across the mosaics. In practice this is not
always true due to the dither pattern used to remove the inter-chip
gaps in the CCD mosaic leading to some small regions with increased
noise when compared to the mosaic as a whole; however, as each
pointing is typically comprised of several individual exposures, the
approximation to a constant noise level is good away from the edges of
the images. We exclude the $U$-band data in constructing the detection
image, as the $BVRIKs$ images contain all the objects detected in $U$,
and the addition of the $U$-band data typically degrades the quality
of the $\chi^2$ detection image.

\subsection{Photometric measurements}
\label{sec:phot-meas}

To construct the photometric catalogs, we run \sex\ using the $\chi^2$
images as the detection images and measure from each of $UBVRIKs$
images with the modal sky level added back, filtering using a Gaussian
matched to the seeing FWHM. Objects are included in the
$\chi^2$-image-detected output catalogs if, in the $\chi^2$ image,
they contain 4 or more contiguous pixels above a pixel detection
threshold at which the distribution of ``sky'' and ``object'' pixels
cross \citep[see][]{1999AJ....117...68S}. This threshold is somewhat
conservative, minimising the number of false detections. Objects
containing bad-pixels (e.g. saturated stars and bleed trails) in
either the detections or measurement images are flagged in the output
catalogs. We measure magnitudes in \citet{1980ApJS...43..305K}-like
apertures as well as a variety of aperture magnitudes up to
$7\farcs5$; where the measured magnitude in a given image is below the
5-$\sigma$ detection threshold, it is flagged by adding 100 to the
measured value. Where a magnitude cannot be measured as the net flux
is negative, the magnitude given is replaced with the formal
5-$\sigma$ limit for that image and is flagged as an upper limit (this
value is more appropriate than, say, the 80\% completeness magnitude
as we know the position at which the object would be if it could be
detected). We set a minimum Kron radius equal to the seeing so that
magnitudes are not measured in unphysically small apertures.

Due to the increased field of view of the $U$-band data, we also
construct catalogs detected in the $U$-band (and measured from the
$UBVRIKs$ as before). A $Ks$-band selected catalog containing only
seeing-matched $RIKs$ magnitudes (and excluding the $UBV$ data with
poorer seeing) and with less conservative detection parameters is also
generated for the purpose of studying the population of objects with
very red colors (Afonso et al., in prep.).

\begin{figure}
  \centering
  \plotone{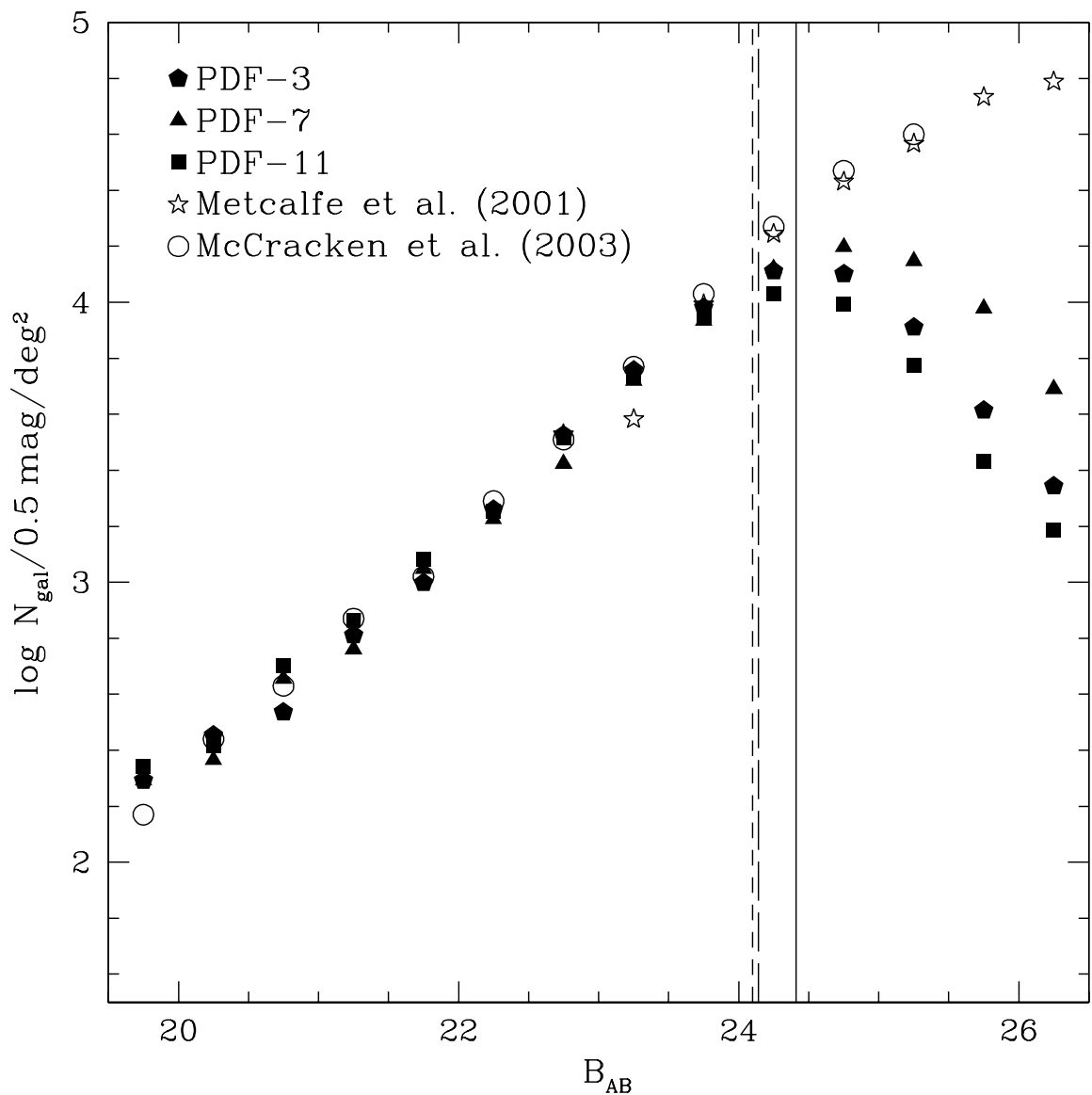}
  \caption{
    The $B_{\mathrm{AB}}$ number counts (derived from \sex\ 
    \textsc{magauto} magnitudes) for the 3 central pointings in our
    optical data, together with the $B$-band counts of
    \citet{2001MNRAS.323..795M} and \citet{2003astro.ph..6254M}. The
    vertical lines show the 80\% completeness limits for PDS-7 (solid
    line), PDS-3 (long-dashed line) and PDS-11 (short-dashed line).
    \label{fig:number-counts}}
\end{figure}

We compare the number counts of our $\chi^2$-detected catalogs in the
$B$-band with published values from deep optical surveys as a
consistency check on the counts derived from this dataset, and to
confirm that the completeness limits that we have derived using
simulations (presented in Table~\ref{tab:completeness}) are consistent
with the incompleteness limits in our actual data.
Figure~\ref{fig:number-counts} shows our $B$-band counts compared with
the deeper counts of \citet{2001MNRAS.323..795M} and
\citet{2003astro.ph..6254M}. Firstly, we see an excellent agreement in
the counts below $B_{\mathrm{AB}}=24$, to which magnitude we can be
reasonably confident our imaging is complete. Secondly, the figure
illustrates that the completeness limits we derived in our simulations
correspond well to the 50 and 80\% limits in our actual data.

\subsection{Radio-optical-spectroscopic identifications}
\label{sec:optical-ids}

We identify optical counterparts to the radio data using the final
$\chi^2$-detected catalogs for the central three pointings, and
$U$-band catalogs for the areas not observed in $BVRIKs$. There are
several techniques which could be used to cross-correlate the radio
and optical catalogs. One statistically robust method is the
likelihood ratio method of \citet{1992MNRAS.259..413S} \citep[but see
also,
e.g.,][]{2001PASP..113...10M,2000ApJS..131..335R,1986MNRAS.223..279W}.
After implementing this method, including the use of an iterative
estimate for $Q$, the prior probability of a counterpart existing with
a particular magnitude, it was found that $<1\%$ of the potential
optical counterpart galaxies differed from those found using a
straightforward nearest-neighbour matching. This is not unreasonable
given the relative sensitivities and positional accuracies of the
1.4\,GHz and optical catalogs. When the few discrepant counterpart
candidates were visually inspected they were all found to be optically
faint galaxies very close to optically bright systems, where the radio
position most closely matched the faint galaxy. In all such cases the
most visually convincing counterpart is the fainter system. Given this
result, and combined with the additional prior information required by
the likelihood ratio method, (some of which, like $Q$, is not
explicitly known), and the simplicity of the nearest-neighbour
matching, we chose to use the nearest-neighbour criterion to identify
the best optical counterpart candidates \citep{2002MNRAS.330..241M}.

We begin, then, by identifying all optical objects within $4''$ of a
radio source. We visually inspect each of these, to further strengthen
the reliability of the final optical counterpart catalog, by creating
``postage" stamps in each optical bandpass for every counterpart, and
overlaying each stamp (including the $\chi^2$ detection image) with
the contours from the radio image. Each of these was visually
inspected to ensure that the optical detection is real and a likely
optical counterpart has been identified. Multiple object counterpart
cases, where more than one object lies within $4''$, are flagged.

A detailed breakdown of the number of radio sources observed and the
number of those which were detected now follows, and is also
summarised in Table~\ref{tab:radiomatches}. Of the 2148 radio
sources available after merging the two catalogs described in
\citet{2003AJ....125..465H}, 839 are located in the three optical pointings
3, 7 and 11, in regions defined as ``good" by the image bad pixel
masks (i.e. 839 radio objects could have been detected in optical
$BVRI$ imaging of this survey). Of these, 827 lie in ``good" (i.e., not masked)
regions of the CTIO $U$-band data. A further 236 are located in the CTIO
$U$-band imaging not overlapped by the $BVRI$ images. Another 256 radio
sources lie within the two pointings (2 and 6) covered solely by the
ESO $U$-band data. Finally 111 radio objects are
located within the smaller area defined by the $Ks$-band mosaic.

To summarise the counterpart matching: 673/839 (79\%) of the radio
detections have candidate optical counterparts in the $\chi^2$ images;
of these 639 (76\%) are detected at $\ge 5\sigma$ in one or more
filters. Taking other bandpasses individually: 569/1063 (54\%) in the
CTIO $U$-band and 53/256 (21\%) in the (shallower) ESO $U$-band have
candidate counterparts detected at $\ge 5\sigma$ significance; 91/111
(82\%) of the radio detections are detected in $Ks$ at $\ge 5\sigma$.
In total, across all the datasets 1331 radio objects are observed, with
candidate optical counterparts identified for 778 (58\%). Again, these
figures are summarised in Table~\ref{tab:radiomatches}. The final
catalog of all 778 detected radio/optical candidate counterparts can
be found in Table~\ref{tab:catalog}.

\begin{figure}
  \centering
  \plottwo{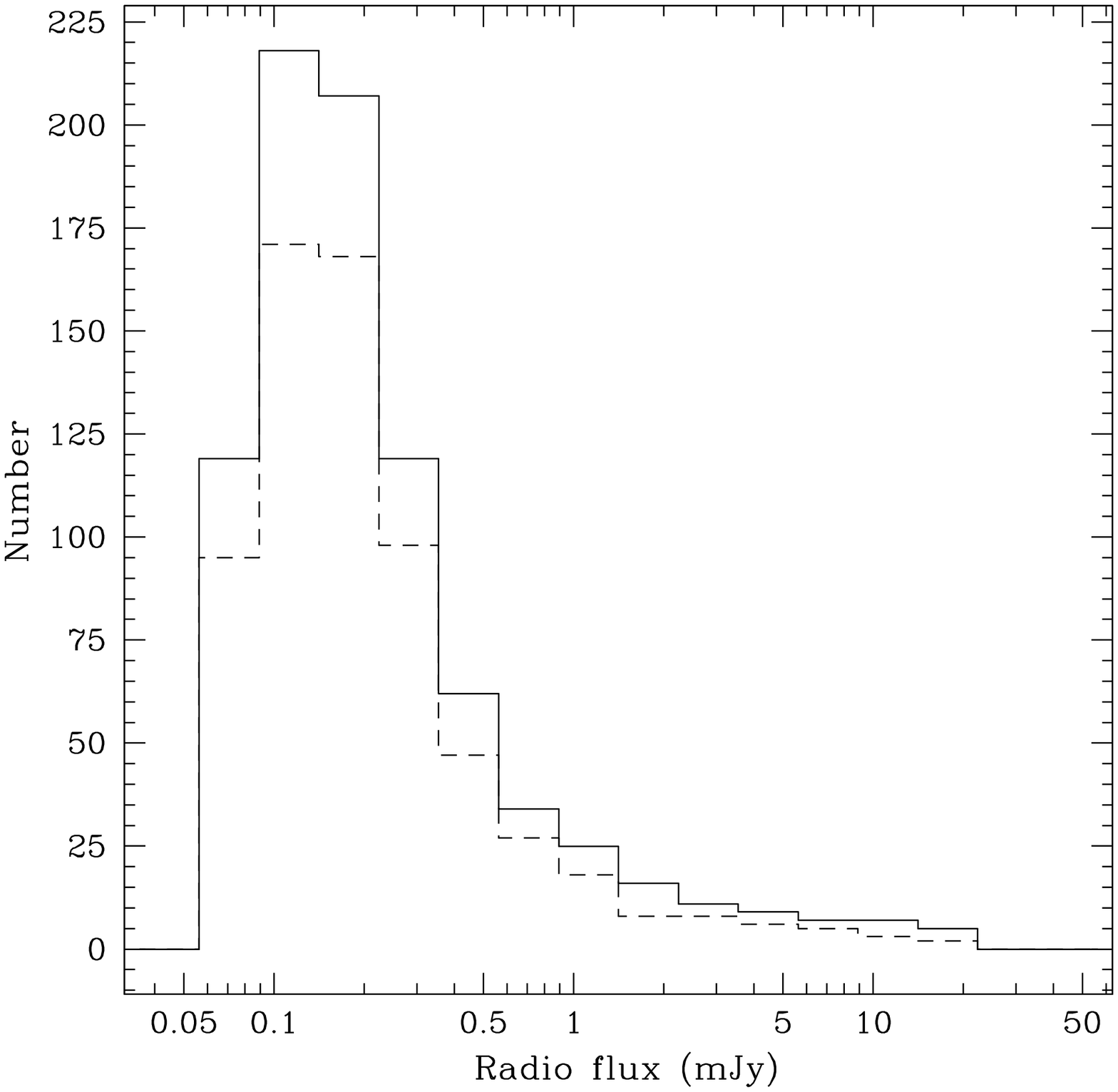}{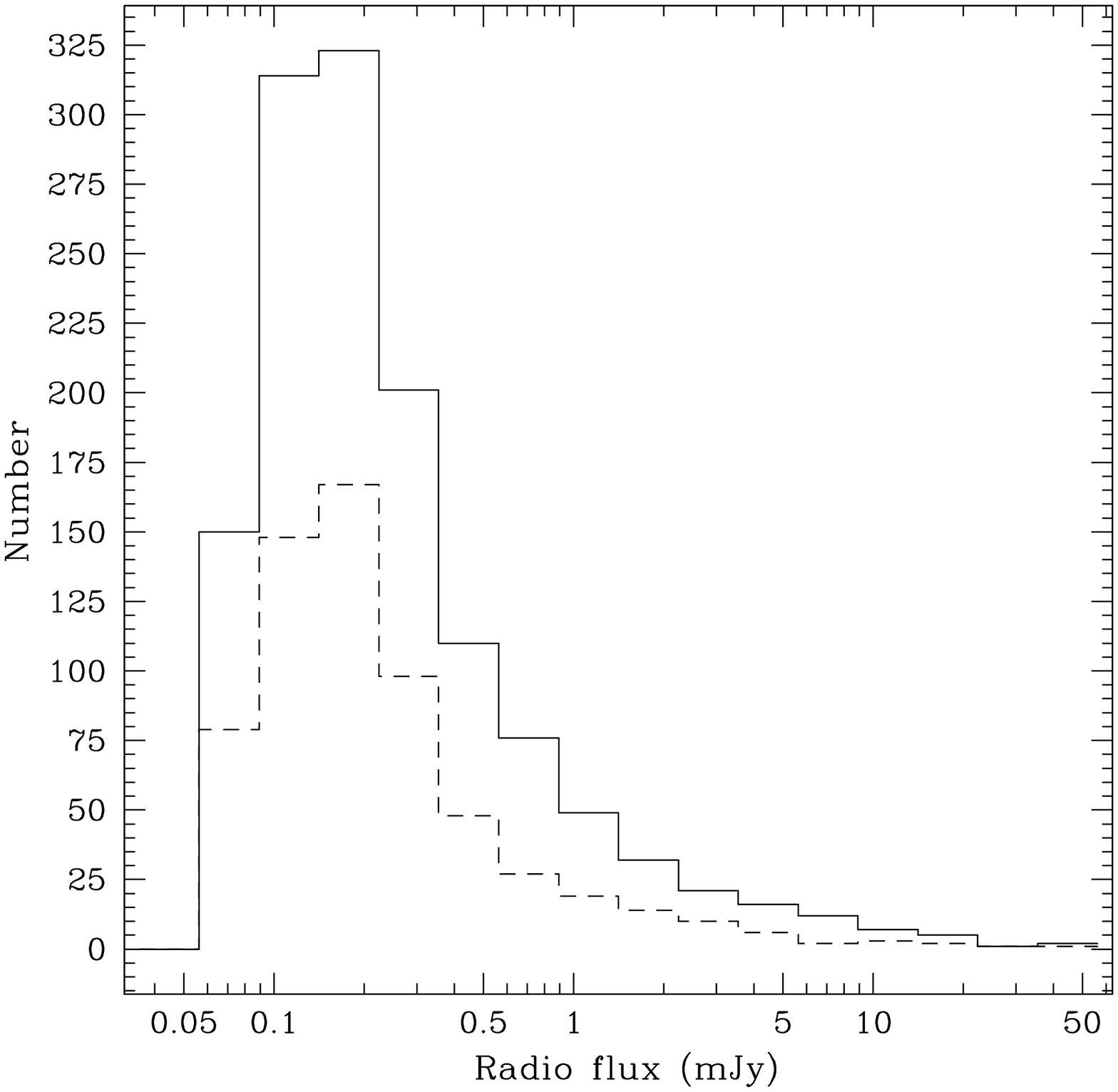}
  \caption{
    The distribution of radio/optical (left) and radio/$U$-band
    (right) candidate matches as a function of radio flux. The solid
    lines show the flux distribution of all radio sources, the dashed
    lines show the flux distribution of the candidate counterparts.
    \label{fig:detections_fn_radio}}
\end{figure}

We show the distribution of our radio galaxies with candidate optical
or $U$-band counterparts as a function of the radio flux in
Figure~\ref{fig:detections_fn_radio}, for those radio galaxies which
were covered by our optical imaging.  We see detection rates of
$>80$\% even in the faintest radio-flux bins, and virtually 100\% for
the brightest radio detections, demonstrating the effectiveness of
using a $\chi^2$-detected catalog. The detection rate in the $U$-band
is lower (though still $>50$\% in all radio-flux bins).

The final step is to cross-correlate the radio-optical catalog with
the spectroscopic observations of the PDS carried our using the 2dF
instrument on the AAT. Full details of these spectroscopic
observations can be found in \citet{1999MNRAS.306..708G} and Afonso et
al. (in prep.). These spectroscopic candidates were selected using
shallower optical observations than those presented in this paper,
effectively limited at $R\sim 22.5$, and hence only the brighter
observations have spectroscopic information available. 237 of our
radio/optical matches have spectral information, also listed in
Table~\ref{tab:catalog} where available.

\section{Photometric redshift analysis}
\label{sec:photo-z}

For the objects in this sample without a spectroscopic redshift, we
can in principle use our extensive $UBVRIKs$ dataset to estimate
photometric redshifts.

\begin{figure}
\epsscale{0.75}
\centerline{\rotatebox{-90}{\plotone{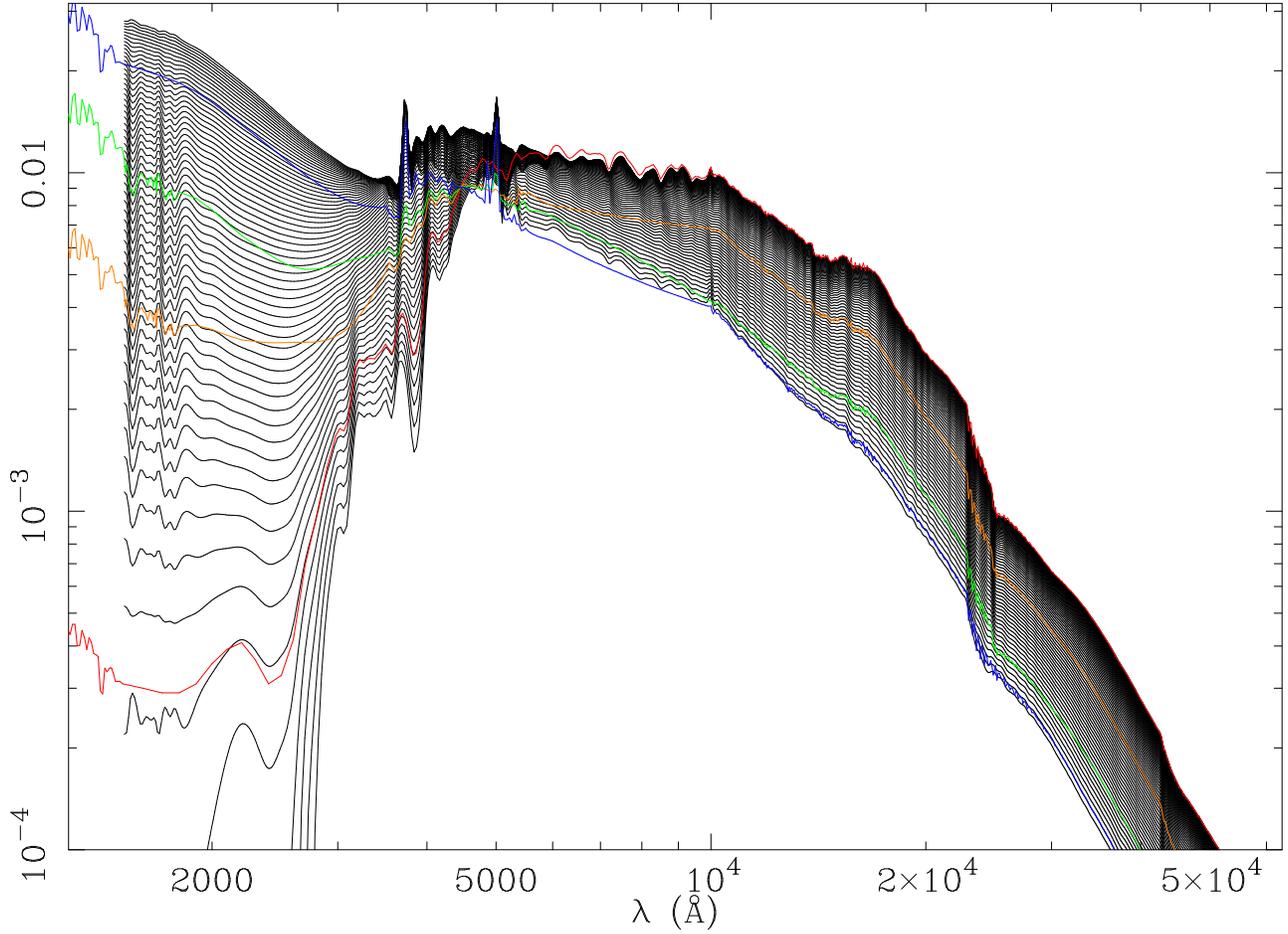}}}
  \caption{The 61 subspace filtered SEDs shown with the 4 original CWW SED
    templates overlaid. Red: E-type (0.0); Orange: Sbc-type (0.33);
    Green: Scd-type (0.67); Blue: Im-type (1.0).
 \label{fig:seds}}
\epsscale{1.0}
\end{figure}

For our analysis, we use the photometric redshift code of
\citet{1995AJ....110.2655C} \citep[see also][]{2000AJ....120.1588B}.
We use SED templates based on those of \citet{1980ApJS...43..393C}
(hereafter CWW), providing four standard SEDs (E/S0, Sbc, Scd, Im),
extended in the UV and IR wavelength regions using the GISSEL98 code
\citep[e.g.][]{1993ApJ...405..538B}. These are ``observed'' SEDs, and
implicitly contain an amount of dust which varies between the
different galaxy classes. Rather than using these four SEDs directly,
we use the method of subspace filtering \citep{2000AJ....120.1588B} to
provide a large number (61) of smoothly interpolated SEDs based on the
reference CWW SEDs.  This supports more realistic type estimates for
most of the galaxies.  These new SEDs are shown in
Figure~\ref{fig:seds}. Our revised type-classification numbering runs
from -0.1 to 1.1, with the standard CWW SEDs lying at positions 0.0
(E), 0.33 (Sbc), 0.66 (Scd), and 1.0 (Im).

We allow the possible photometric redshifts to range from 0.0 to 1.3
and also apply a very weak prior constraining the absolute magnitude
range of the galaxies to $-29<M_B<-16$, which has the effect of
eliminating photometric redshift fits that produce unphysically
extreme luminosities. Allowing maximum redshifts greater than 1.3
introduced a large number of false photometric redshifts close to the
maximum value, as at $z\sim1.4$ the 4000\,\AA\ break spectral feature
moves out of the band-passes of our optical imaging and some
low-redshift objects can then be spuriously fit with photometric
redshifts $z>1.4$. Limiting the redshift range to 1.3 eliminated this
problem.  While this strategy excludes the possibility of accurately
identifying the highest redshift counterparts to faint radio sources,
previous studies covering smaller survey areas
\citep[e.g.][]{2000PASP..112.1001R}, and basing classifications on
radio spectral index and optical counterpart \textit{HST} imaging,
show that $\simeq60$\% of the radio population to $I_{\mathrm{AB}}<25$
is composed of star-forming galaxies lying at $0<z<1.3$, 20\%
comprising AGN/QSOs, with the remainder classed as ambiguous. Hence
almost all of the star-forming systems in the 1.4\,GHz catalog (and
which are the primary focus of our companion papers) will actually lie
at $z<1.3$, and only a small fraction of objects (mostly QSOs and
AGNs), will be lost. (This class of object is discussed further
below).

\begin{figure}
\epsscale{0.75}
\centerline{\rotatebox{-90}{\plotone{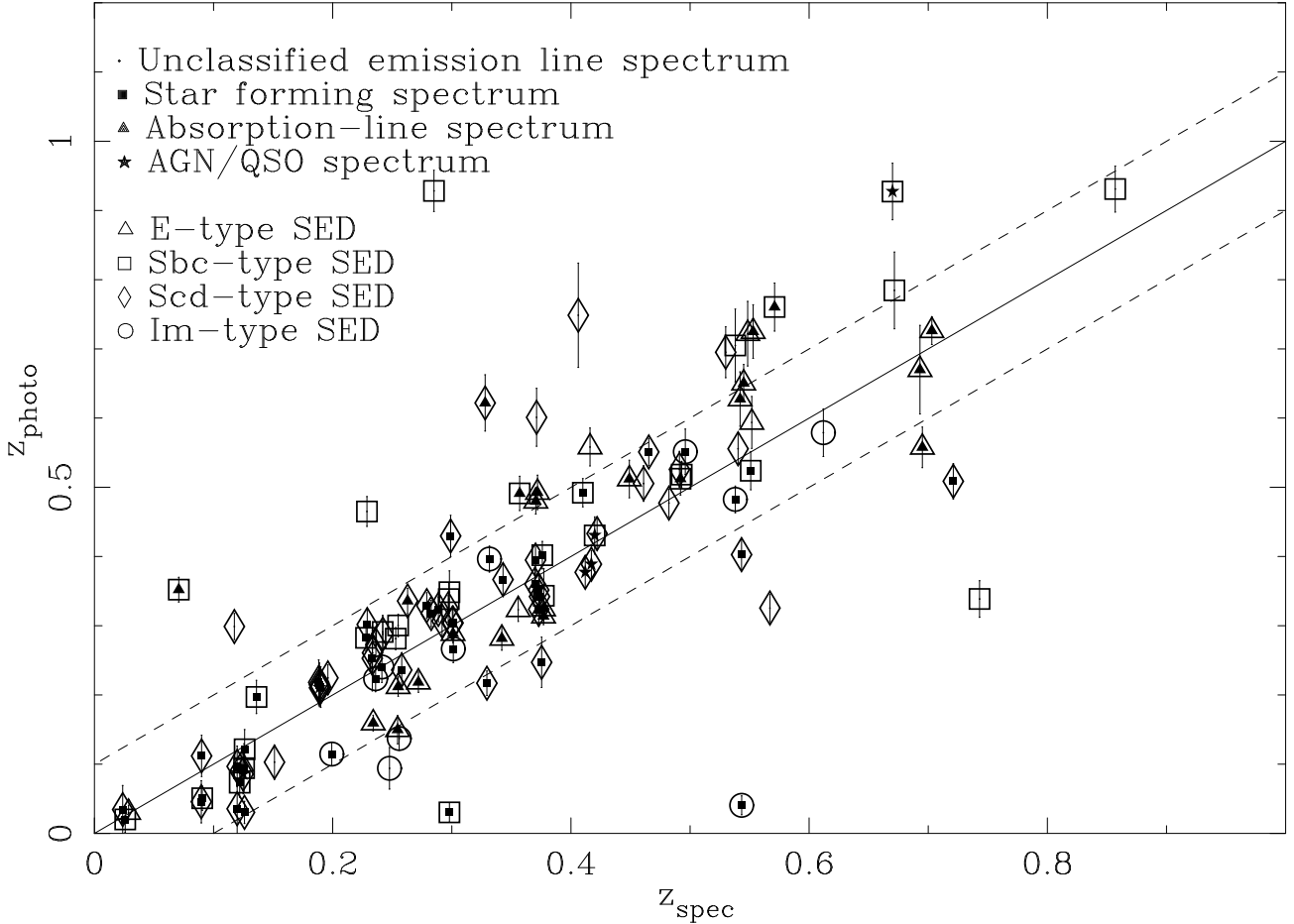}}}
  \caption{A comparison of spectroscopic and photometric redshifts for
    the spectroscopically-observed sub-sample. The points and filled
    symbols refer to the spectroscopic classification of the galaxies
    \citep{1999MNRAS.306..708G}, while the open symbols refer to the
    best-fitting SED from the photometric redshift estimation. The
    open symbols should be treated as a rough guide only, since they
    group the 61 subspace filtered SEDs into the closest CWW-type
    classification, with SEDs from -0.10--0.20 called E-type,
    0.21--0.50 Sbc-type, 0.51--0.80 Scd-type and 0.81--1.10 Im-type.
    Note that this figure shows redshift in linear units, with the
    dashed lines indicating offsets of $\pm0.1$ in $z_{\rm photo}$
    about the one-to-one line.
 \label{fig:photo-compare}}
\epsscale{1.0}
\end{figure}

We first examine the spectroscopically observed galaxies by comparing
the spectroscopic and photometric redshift.
Figure~\ref{fig:photo-compare} shows this for a sample of 116 galaxies
with measured $UBVRI$ magnitudes. The filled symbols here (including
points) indicate the spectroscopic classification, while the open
symbols give an estimate of the best-fitting SED template type (after
``binning" the 61 subspace filtered templates into four bins, based
approximately on the closest CWW-type template). As well as providing
reasonable redshift estimates, the best-fitting SED is also a good
indicator of the galaxy type, in the sense that spectroscopic
absorption-line systems are mostly well-fit by early-type SEDs, while
star-forming systems are mostly well-fit by late-type SEDs.

\begin{figure}
\epsscale{0.75}
\centerline{\rotatebox{-90}{\plotone{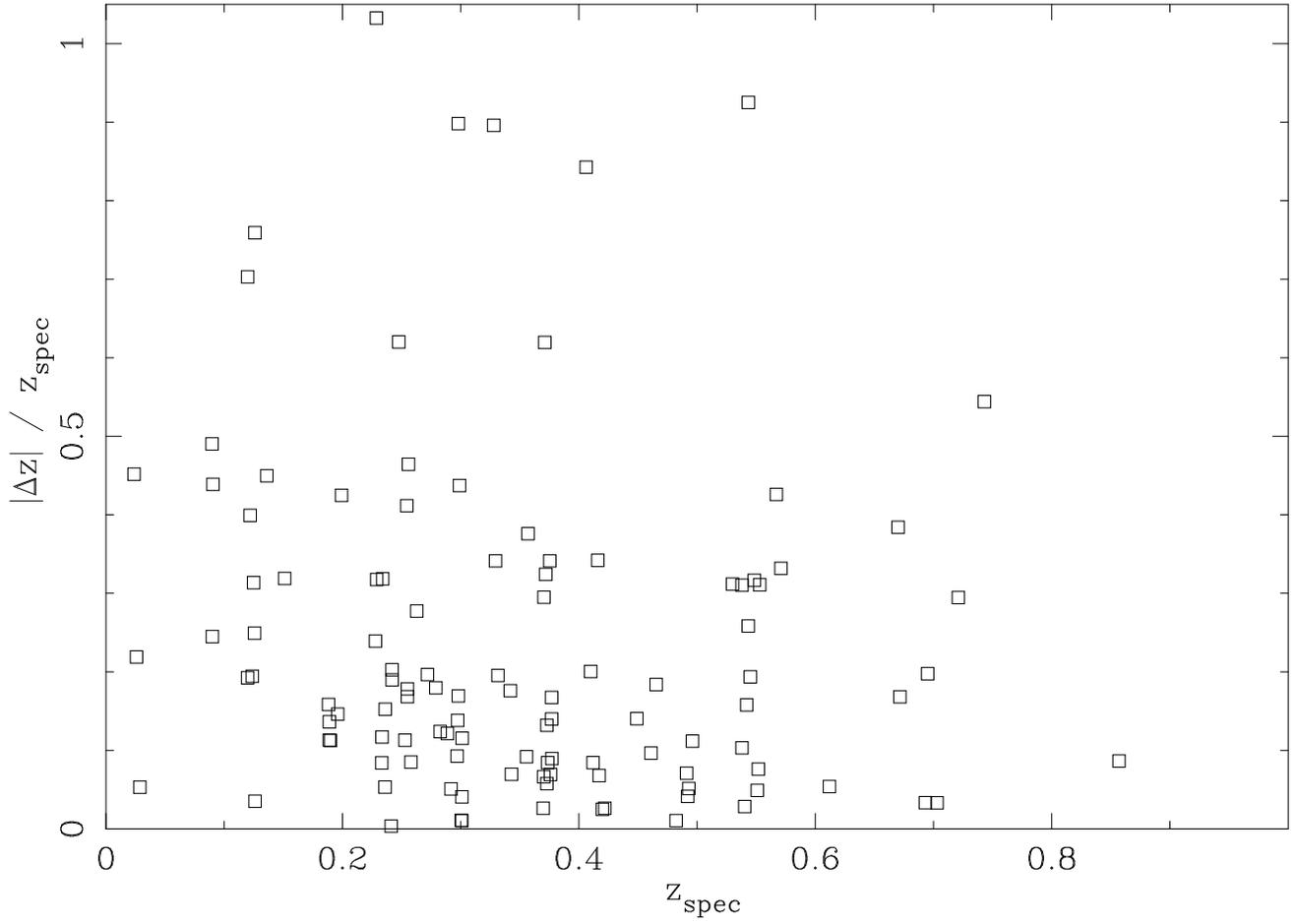}}}
  \caption{Distribution of $|\Delta z|/z_{\rm spec}$ with spectroscopic
redshift.
 \label{fig:photo-err1}}
\epsscale{1.0}
\end{figure}

\begin{figure}
\epsscale{0.75}
\centerline{\rotatebox{-90}{\plotone{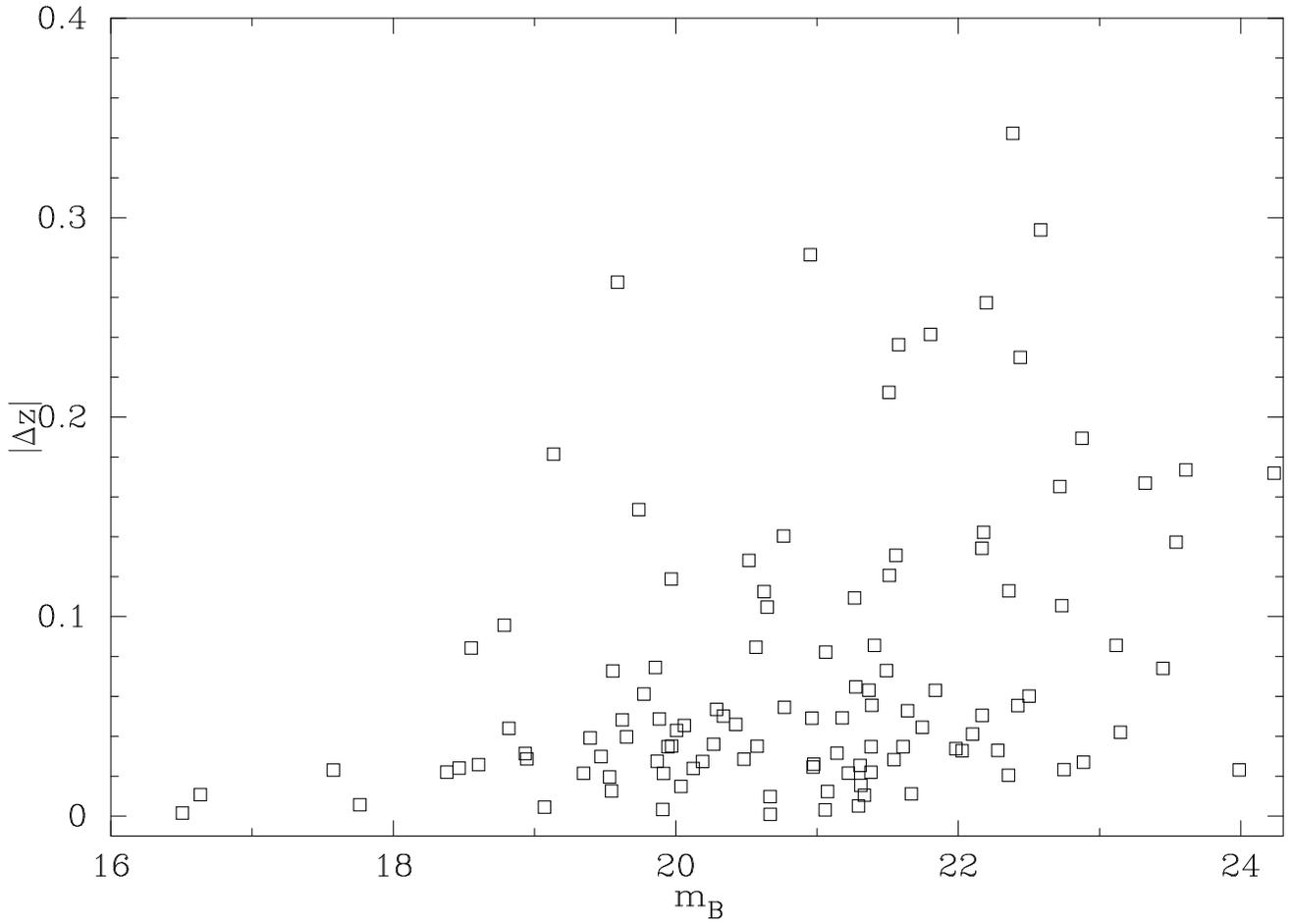}}}
  \caption{Distribution of $|\Delta z|$ with B-band magnitude,
showing that (as expected) photometric redshift uncertainties increase for fainter systems.
 \label{fig:photo-err2}}
\epsscale{1.0}
\end{figure}

The reliability of the photometric redshifts can be characterised in
several ways. The rms of $|\Delta z| = |z_{\rm photo}-z_{\rm spec}|$
is 0.1 for these 116 galaxies, over the redshift range $0<z<1$. The
rms of $|\log[(1+z_{\rm photo})/(1+z_{\rm spec})]|$ is 0.028 (implying
a typical uncertainty of $7\%$ in $1+z_{\rm photo}$). The rms of
$|\Delta z|/(1+z_{\rm spec})$ is 0.065. This level of reliability in
the photometric redshifts compares favourably with that of other
analyses based on data with similar photometric uncertainties
\citep{1998AJ....115.1418H,1999ApJ...513...34F,2001ApJS..135...41F,2003MNRAS.345..819R}.
The relative error, $|\Delta z|/z_{\rm spec}$ is shown as a function
of $z_{\rm spec}$ in Figure~\ref{fig:photo-err1}, where it is clear
that the roughly constant rms uncertainty in $z_{\rm photo}$ with
redshift translates into a smaller relative error at higher redshift.
The error $|\Delta z|$ is shown as a function of apparent $B$-band
magnitude, $m_B$, in Figure~\ref{fig:photo-err2}, where it is clear
that the larger absolute uncertainties in $z_{\rm photo}$ tend to
occur for fainter galaxies due to larger photometric uncertainties.

The addition of $Ks$-band magnitudes to the $z_{\rm photo}$ estimate is
likely to improve the reliability somewhat, though this was harder to
quantify for our sample, with only 37 objects having both
spectroscopic redshifts and $UBVRIKs$ photometry. The measured
uncertainties from this small sample were similar to those when using
only $UBVRI$ photometry, and a fuller analysis must await further
near-IR imaging. Further experimentation established that in the
absence of $R$-band photometry (i.e. using only $UBVI$ magnitudes),
$z_{\mathrm{photo}}$ estimates could still be obtained with a
reliability only marginally worse than above. However, the presence of
an $R-I$ color term to correctly calibrate our $I$ photometry to the
standard system required that photometric redshifts could only be
estimated for objects with both $R$ and $I$ photometry. Eliminating
other filters significantly degraded the quality of the photometric
redshifts, and as a result we calculate $z_{\rm photo}$ only for
sources having measurements including at least $UBVRI$.

\begin{figure}
  \plottwo{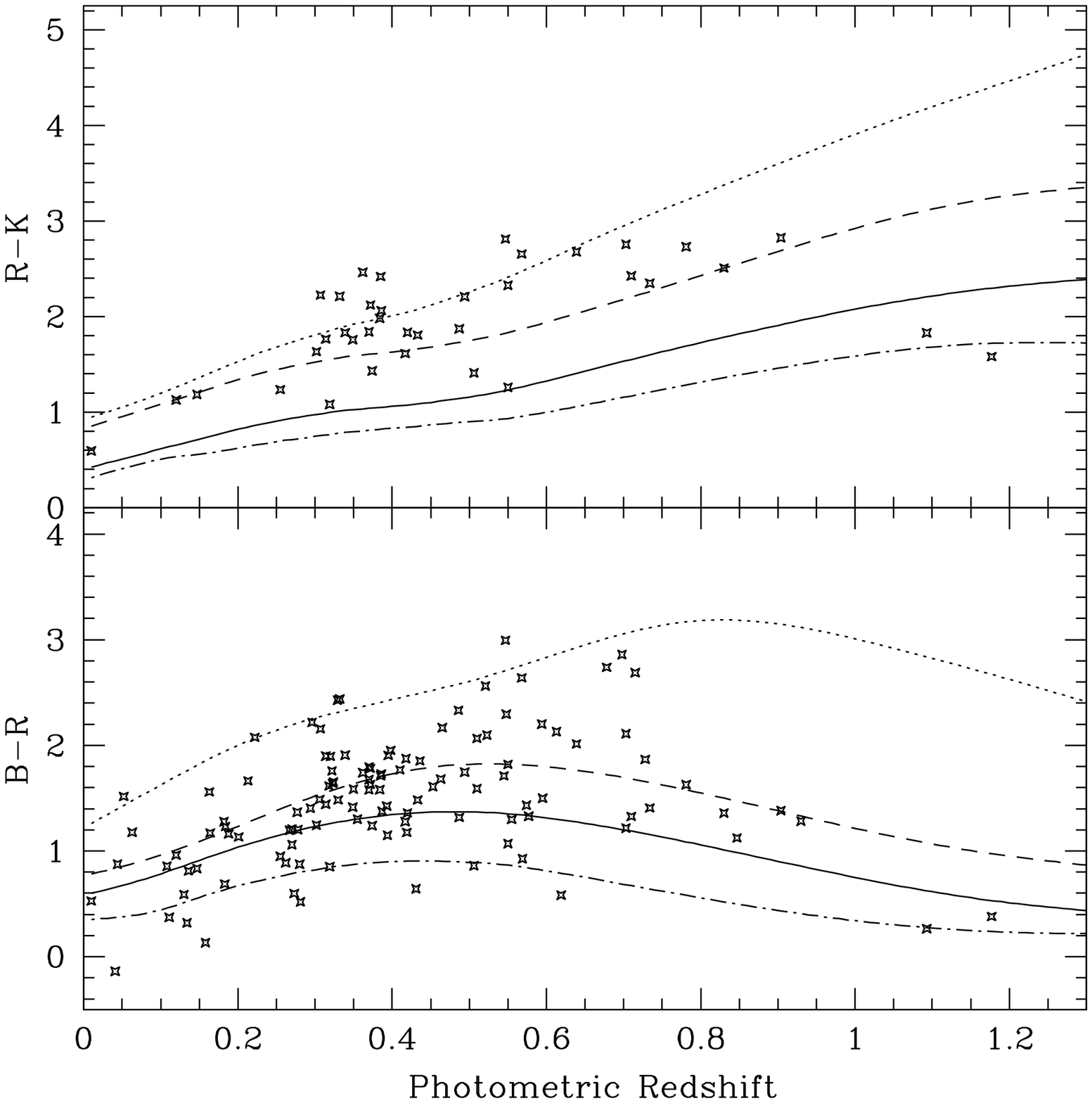}{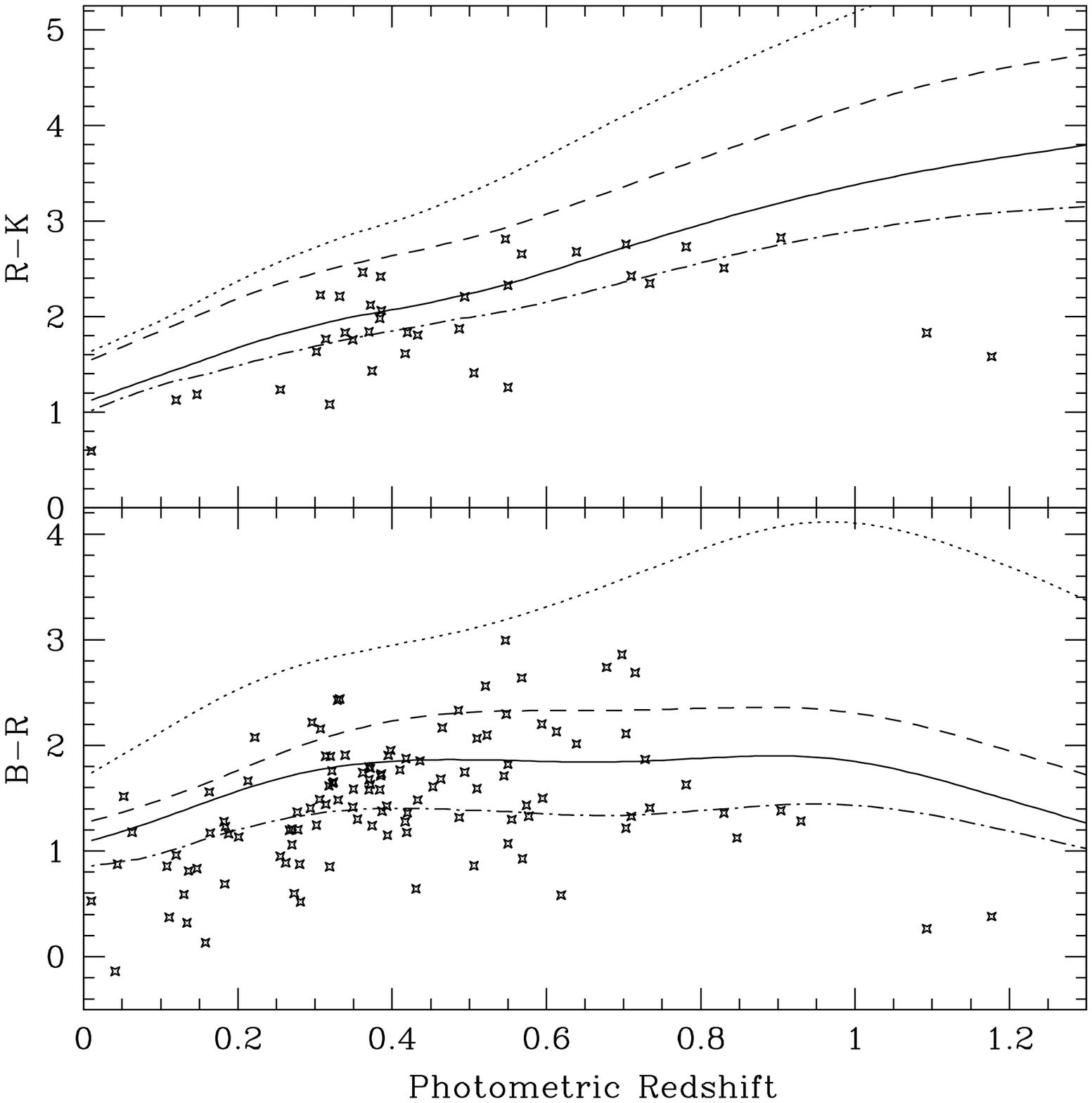}
  \caption{
    $R_{\rm AB}-Ks_{\rm AB}$ (UPPER) and $B_{\rm AB}-R_{\rm AB}$
    (LOWER) as a function of redshift, compared with the colors
    corresponding to the SEDs used in the photometric redshift
    estimation. Both panels show SED colors for the 4 CWW-type SEDs
    used (upper SED to lower SED these are E: dotted, Sbc: dashed,
    Scd: solid, Im: dot-dashed). The curves in the right-hand panel
    have additional reddening ($A_{\rm V}=1.0$) compared to the
    templates used for deriving photometric redshifts.
 \label{fig:color_redshift}}
\end{figure}

A brief investigation of the reliability of the templates chosen is
indicated in Figure~\ref{fig:color_redshift}, showing $(R-Ks)_{\rm
  AB}$ and $(B-R)_{\rm AB}$ against photometric redshift. The observed
colors are compared with loci indicating the color-redshift relation
based on the CWW SEDs used, and also compared with SEDs containing
additional reddening \citep{1999MNRAS.306..708G}.  It is clear that
little or no additional reddening needs to be invoked out to $z\approx
1$ to produce the observed galaxy colors for this sample.

It should be noted that the photometric redshift estimates so far
described neglect the AGN or QSO population, since no spectral
templates representing these systems have been added to the CWW SED
set used. Adding such templates from a variety of sources was
extensively explored, in an attempt to account simultaneously for both
the ``normal" galaxies and the AGN/QSO population.  None of the small
number of QSOs and narrow-line AGN in the spectroscopic sample,
however, were fit better by these templates than by the CWW-style
templates. For the narrow-line AGNs, the CWW-style SEDs still give
rise to acceptable photometric redshifts, with the colors of such
objects being fit well by early-type galaxy SEDs (the points indicated
with filled stars in Figure~\ref{fig:photo-compare}). None of the
three QSOs (at $z>1$) in the spectroscopic sample were well fit by any
of the SEDs explored, and typically were seen to produce erroneous
photometric redshifts of $z_{\rm photo}\approx 0.1$. The implication,
then, of neglecting AGN/QSO SEDs in the photometric redshift
estimation will primarily be the erroneous classification of the small
QSO population as low-redshift galaxies. Further work is clearly
required to identify AGN/QSOs among the faintest optical counterparts
of the PDS radio sources; however as there are few such objects
contributing to the total galaxy sample (at least in our existing
spectroscopic sample) this systematic is likely to be small when
considered in combination with the intrinsic uncertainties attached to
the photometric redshifts themselves.

A final population of galaxies that is overlooked in the current
photometric redshift analysis is that of the extremely red galaxies
(ERGs).  Preliminary estimates suggest that over 400 ERGs (defined in
this case by the color criterion $R-Ks>5$) exist in the NIR imaging
survey, a small fraction of which (about 20 galaxies) are identified
with radio counterparts \citep{2003astro.ph..9147H}.  The current
photometric redshift analysis makes no attempt to classify these red
SED galaxies due to the non-detection of a large fraction of these
objects in one or more of the shortest wavelength passbands, where the
red colors of the SEDs makes them extremely faint. The ERG population
as a whole will be treated in detail in a subsequent paper using SEDs
more appropriate to the presumed ERG population (Afonso et al., in
prep.).

Our final photometric redshift estimates for the full sample of
galaxies containing $UBVRI$ photometry can be found in
Table~\ref{tab:catalog}.

\section{General photometric properties of the sample}
\label{photometric-properties}

In this section we examine the broad photometric properties of the
radio-selected sample.

\subsection{Galaxy types}
\label{galaxy-types}

To correct the observed magnitudes into rest-frame magnitudes we must
apply a $k$-correction \citep[see, for
example,][]{1988ApJ...326....1Y,1997A&AS..122..399P}. The magnitude of
the $k$-corrections for template galaxy spectra using the range of the
filters in this study strongly depends on galaxy type, with early-type
galaxies (ellipticals and S0s) possessing larger $k$-corrections in
the optical passbands than galaxies with flatter spectra. We assign
each object in our survey a $k$-correction by fitting the observed
magnitudes to a set of appropriately redshifted model galaxy templates
of varying type. We use the popular technique of fitting a series of
template SEDs to the broadband colors available for each galaxy and
use the best-fitting SED for the estimation of $k$-corrections in each
filter, linearly interpolating between the SED types. These fits are
combined by weighting inversely by the variance in the observed colors
to obtain a mean best-fitting SED for each galaxy. This process was
performed for the photometric-redshift sample in
Section~\ref{sec:photo-z}; here we repeat the exercise for the
spectroscopic sub-sample. We use the subspace-filtered SEDs of
\citet{1980ApJS...43..393C} as in the photometric redshift analysis.

\begin{figure}
  \centering
  \plotone{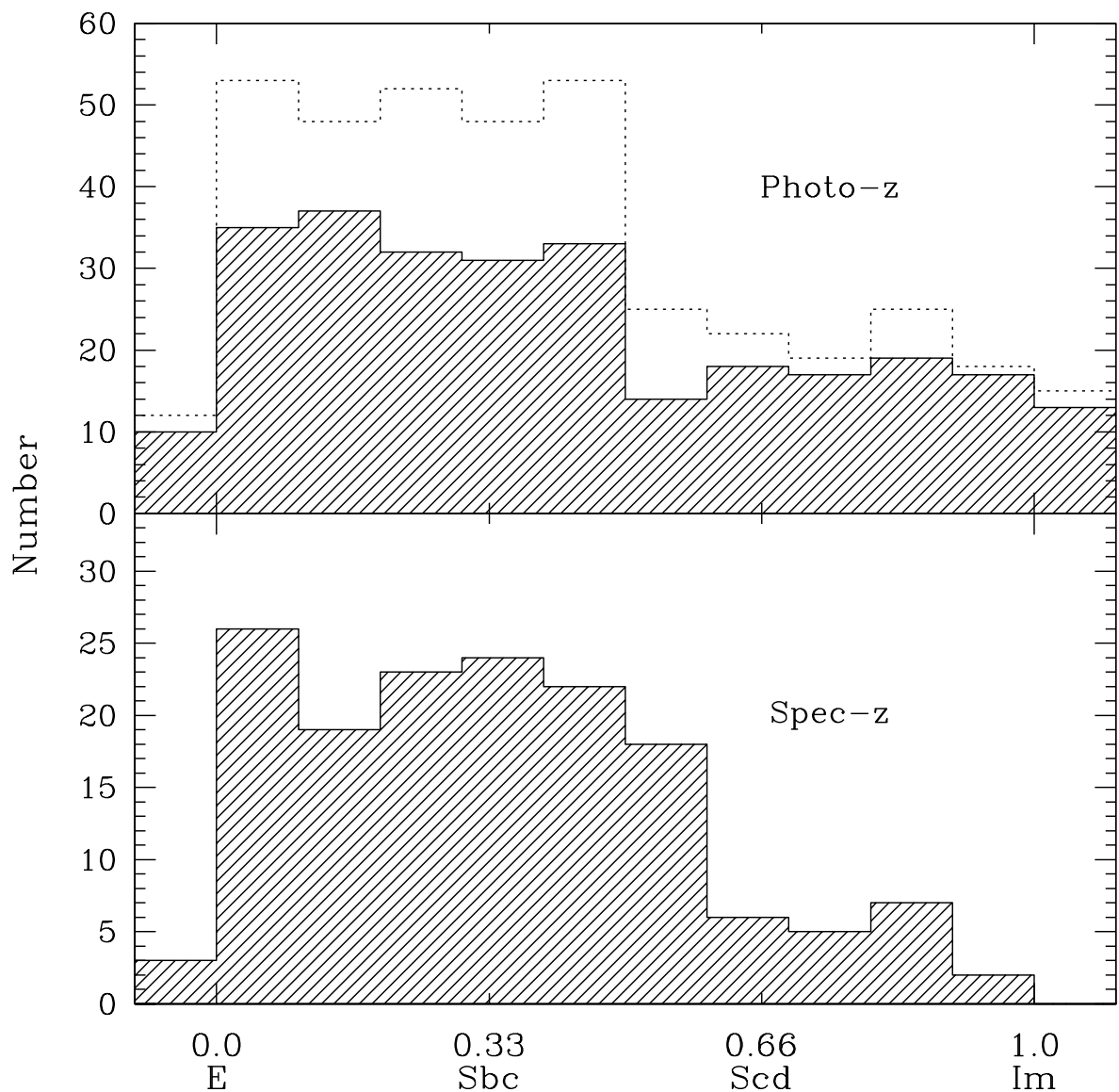}
  \caption{
    The distribution of galaxy types for the sub-sample with
    spectroscopic (lower) and photometric (upper) redshifts calculated
    using the CWW SEDs. The dotted line in the photometric-redshift
    type distribution shows the location of those galaxies with a
    spectroscopic redshift for which a photometric estimate has also
    been estimated. The median type for both samples is approximately
    an Sbc SED. No dust is added to the CWW SEDs in this analysis.
    Galaxies spectroscopically classified as AGN are omitted from
    this analysis.\label{fig:galaxy_types}}
\end{figure}

In Figure~\ref{fig:galaxy_types}, we present the results of the fitting
process, showing the distribution of the spectroscopic and photometric
redshift galaxy types ($T$), with $T=0$ representing the E SED and
$T=1$ the Im SED.  Figure~\ref{fig:galaxy_types} shows the distribution
of galaxy types for the spectroscopic subsample. The mean type is
$T=0.36$ ($\simeq$~Sbc) for the spectroscopic and $T=0.39$ for the
photometric redshift galaxies. The distribution of the
spectroscopic-redshift types shows an absence of blue systems due to
the original $R$-band limited selection of the spectroscopic follow-up
sample. The photometric-redshift distribution is flatter, with no
particular pre-dominance of any one galaxy SED. There is also a good
agreement seen between the derived types for those objects with
photometric redshifts compared to spectroscopic redshifts, as shown by
the dotted line in the plots.

\subsection{Color-color diagrams}
\label{sec:color_color}

\begin{figure}
  \centering
   \plottwo{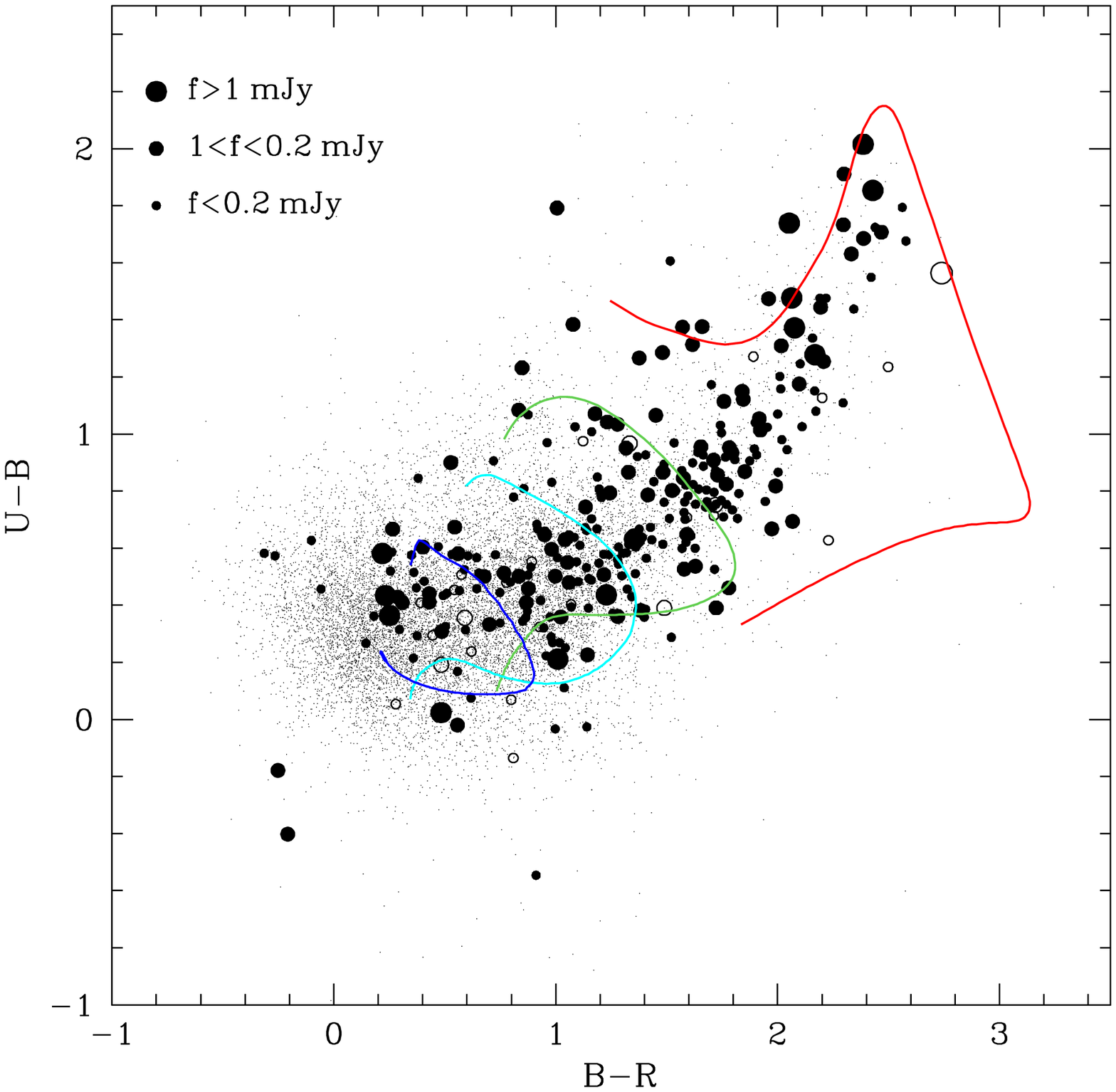}{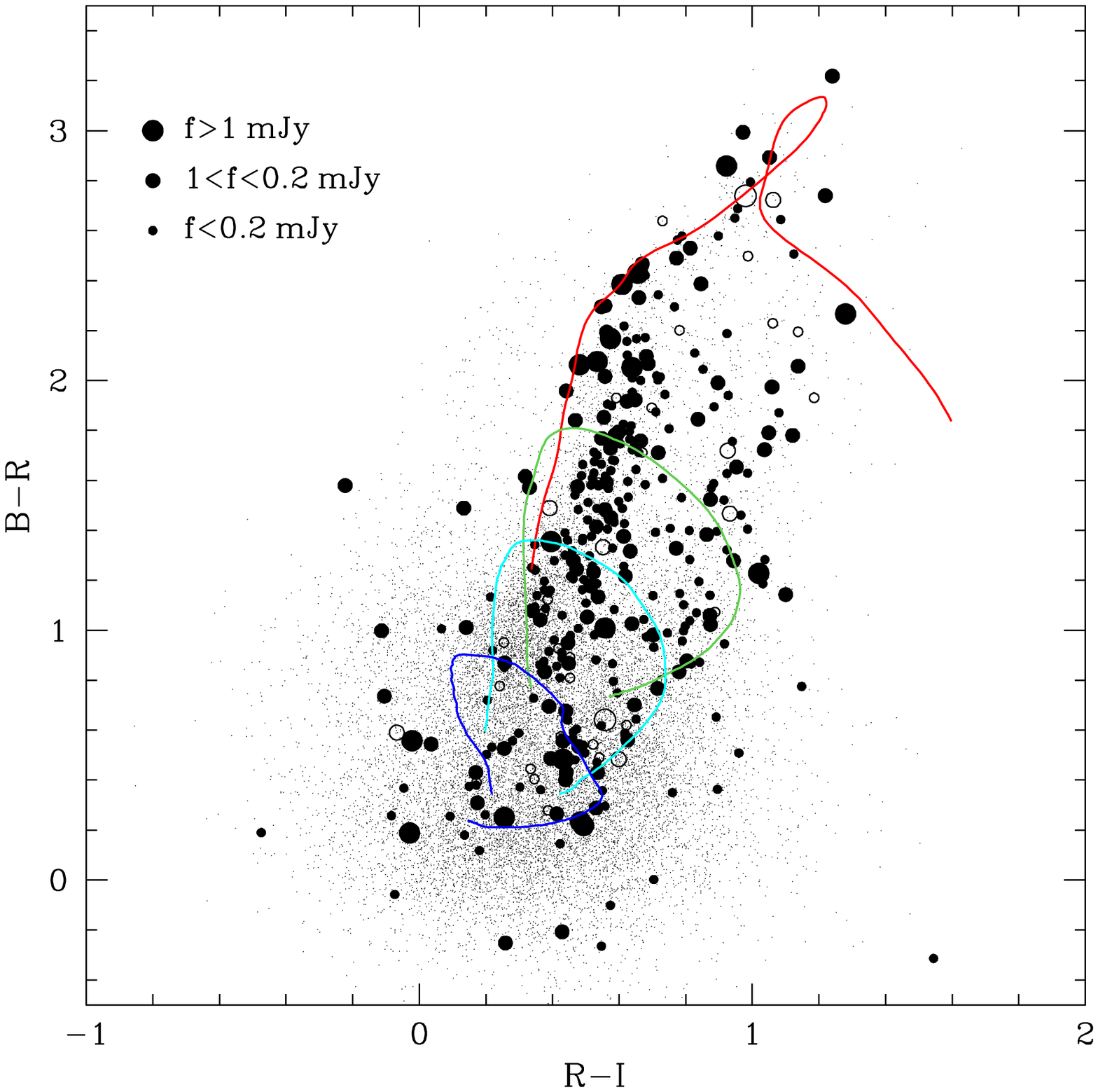}
  \caption{
    Color-color diagrams for the radio population compared to objects
    in the field as a whole. Radio objects are marked as circles where
    the size denotes the radio flux of the source according to the
    key. Filled circles are radio objects with single optical
    counterparts within $3''$, open circles are objects with
    more than one optical counterpart within this radius. The overlaid
    lines represent the position in color-color space of the CWW
    SEDs used to determine the photometric redshift estimates placed
    between a redshift of 0 and 1.3.
\label{fig:color-color}}
\end{figure}

In the absence of spectroscopic information for the optically faint
radio population, color-color diagrams provide a related and
complementary method to that of photometric redshifts to statistically
examine the physical nature of the population, as well as a comparison
with standard galaxy SEDs. We present color-color diagrams, together
with the predictions of the expected colors of optically faint
galaxies based on the CWW SEDs, in Figure~\ref{fig:color-color}. This
confirms that the CWW SEDs are appropriate for exploring the
photometric redshifts, as they appear to adequately span the
color-color space occupied by the catalogued galaxies.  A detailed
analysis of the colors of the radio-detected population will be
explored in a subsequent paper (Sullivan et al., in prep.), also
addressing the star-formation and obscuration properties of the
optically-detected galaxies from the 1.4\,GHz catalog.

\subsection{Distribution by type}
\label{sec:type}
\begin{figure}
\plotone{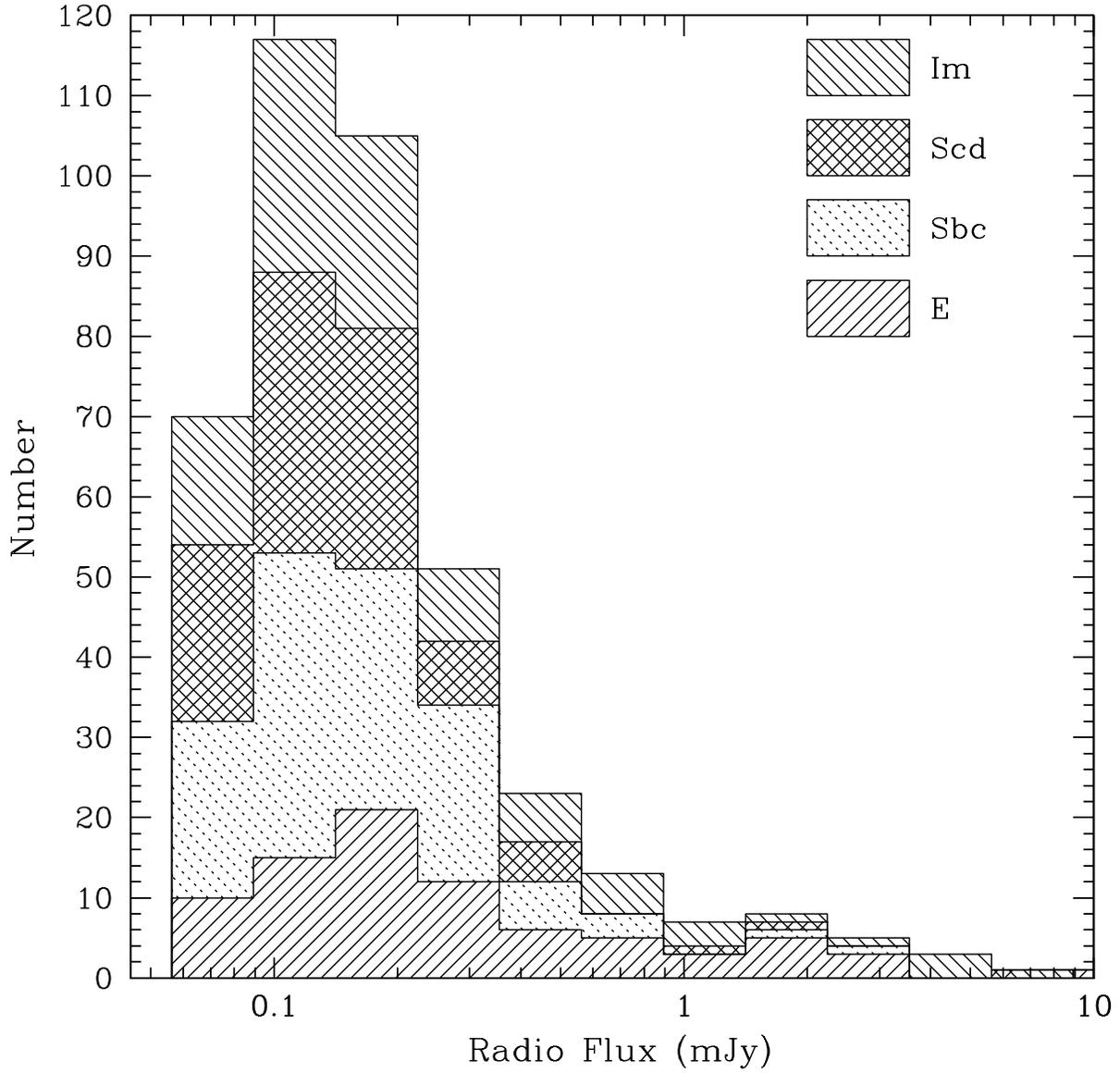}
  \caption{Histogram showing distribution in $S_{1.4}$ by best-fitting
    photometric redshift SED type. The four type bins are as described
    in the caption of Figure~\ref{fig:photo-compare}, with E-type
    galaxies at the bottom of the plot moving through Sb and Sc types
    to Im-type galaxies at the top.
 \label{fig:flxtyp}}
\end{figure}

\begin{figure}
\plotone{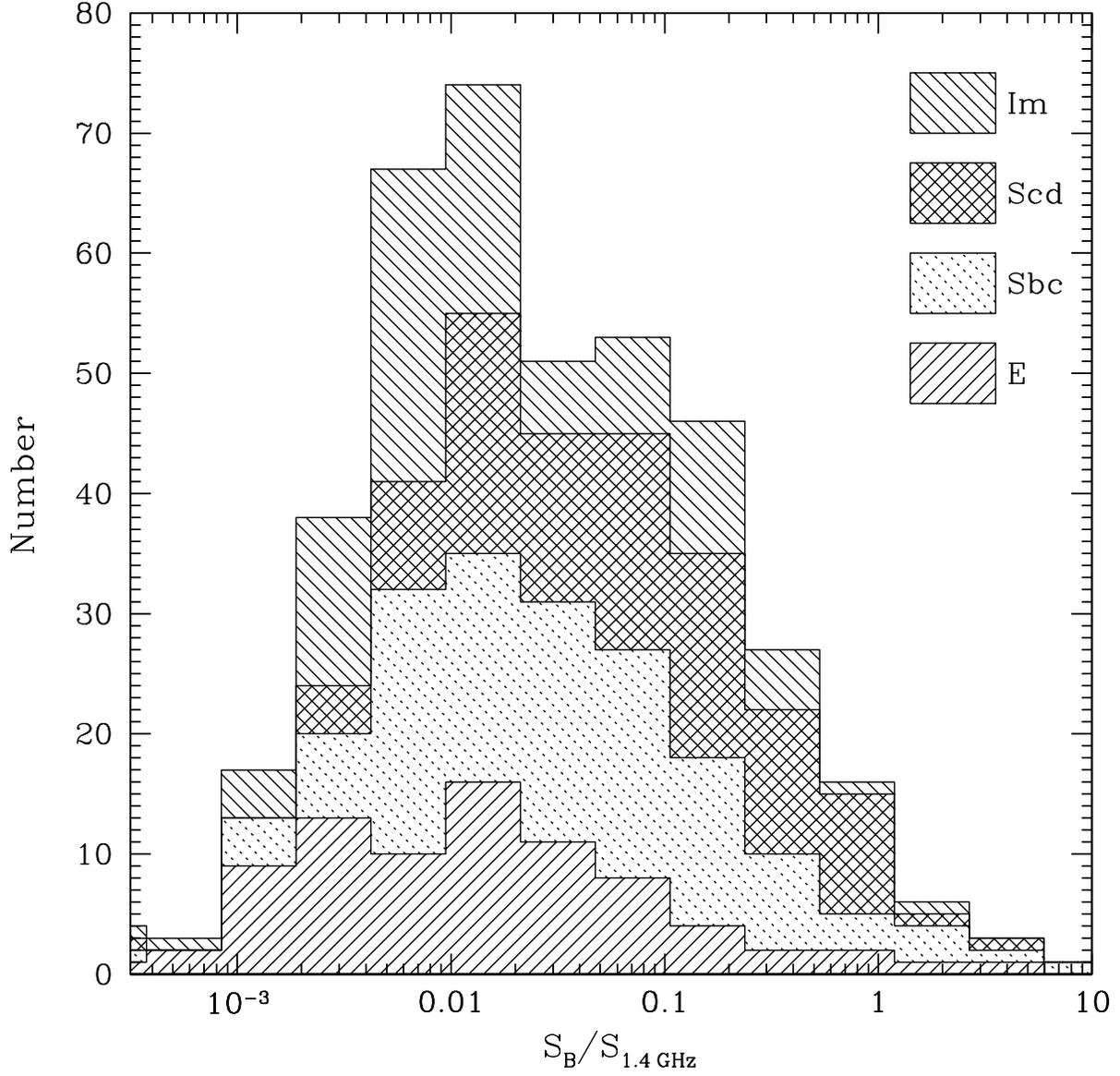}
  \caption{Distributions of optical ($m_B$) to 1.4\,GHz flux ratio,
divided by type. Best-fitting SED types as in previous Figure.
 \label{fig:flxrat}}
\end{figure}

There is a trend in Figure~\ref{fig:color-color} for the lower
1.4\,GHz flux density objects to preferentially lie in the region of
color-color space described best by later-type SEDs. This is explored
more explicitly in Figure~\ref{fig:flxtyp}, showing histograms of
$S_{1.4}$ for four bins in best-fitting SED type, based on the
photometric redshift results.  It is clear that the lower flux
densities have an increasing proportion of later-type SEDs, consistent
with earlier results
\citep[e.g.,][]{1993MNRAS.263...98B,1985ApJ...289..494W}.
Figure~\ref{fig:flxrat} shows histograms of $S_{B}/S_{1.4}$ split by
best-fitting SED type. Systems with higher $S_{B}/S_{1.4}$ ratios are
also seen to be dominated by later type SEDs, illustrating the larger
apparent contribution of late-types at lower 1.4\,GHz flux densities,
as the distribution of $S_{B}$ is fairly uniform over all SED types.
Clearly there are complex selection effects at work which affect these
distributions which will be analysed further in our later papers.

\section{Summary}
\label{sec:summary}

In this paper, we have presented complementary optical $UBVRIKs$
imaging to the Phoenix Deep Survey
\citep[PDS;][]{2003AJ....125..465H}.  The optical $UBVRI$ imaging
covers an area on the sky of $\simeq1$ square degree to $R\sim24.5$,
with a smaller sub-region of near-IR $Ks$ data.  Within this optical
imaging, 778 radio sources can be assigned candidate optical
counterparts. We have explored photometric redshift calculation
techniques and present these estimates for each of the candidate
optical counterparts. These, in combination with the catalogs and data
described above, will be used to support future detailed analyses of
the PDS galaxies.

\acknowledgments The authors would like to thank Andrew Connolly and
David Gilbank for useful discussion.  MS acknowledges support from a
PPARC fellowship. AMH acknowledges support provided by the National
Aeronautics and Space Administration (NASA) through Hubble Fellowship
grant HST-HF-01140.01-A awarded by the Space Telescope Science
Institute (STScI).  JA gratefully acknowledges the support from the
Science and Technology Foundation (FCT, Portugal) through the
fellowship BPD-5535-2001 and the research grant POCTI-FNU-43805-2001.
Based on observations obtained at Cerro Tololo Inter-American
Observatory a division of the National Optical Astronomy
Observatories, which is operated by the Association of Universities
for Research in Astronomy, Inc. under cooperative agreement with the
National Science Foundation. Based on data acquired with the
Wide-Field Imager at the Anglo-Australian Observatory. Based on
observations made with ESO Telescopes at the La Silla Observatory
under programme IDs 66.A-0193 and 67.A-0403.


\clearpage

\clearpage 

\clearpage

\clearpage

\clearpage

\clearpage

\clearpage

\clearpage

\clearpage

\clearpage

\clearpage

\clearpage

\clearpage

\clearpage

\clearpage

\begin{deluxetable}{ccccccccc}
  \tablecaption{Positions of the PDS-optical pointings from Figure~\ref{fig:pds_coverage} and exposure times.\label{tab:optical_observations}}
  \tablewidth{0pt}
  \tabletypesize{\scriptsize}
  \tablehead{\colhead{Pointing} & \colhead{RA (J2000)} & \colhead{DEC (J2000)} & \colhead{$U^a$} & \colhead{$B$} & \colhead{$V$} & \colhead{$R$} & \colhead{$i$} & \colhead{$K$}}
  \startdata
  2  & 01:14:32 & -45:10:20 & 8$\times$420s & \nodata & \nodata & \nodata & \nodata & \nodata\\
  3  & 01:11:29 & -45:10:20 & 5$\times$1200s & 5$\times$480s & 10$\times$240s & 8$\times$240s & 8$\times$480s & \nodata\\
  6  & 01:14:32 & -45:45:30 & 13$\times$240s & \nodata & \nodata & \nodata & \nodata & \nodata\\
  7  & 01:11:29 & -45:45:30 & 6$\times$1200s & 10$\times$480s & 7$\times$300s & 10$\times$480s & 15$\times$480s & 2700s-10800s \\
  11 & 01:11:29 & -46:20:20 & 6$\times$1200s & 5$\times$480s & 5$\times$480s & \nodata & 2$\times$300s & \nodata\\
  \enddata
\tablenotetext{a}{Pointings 3,7,11 were observed with Mosaic-II on the CTIO 4m; pointings 2,6 with WFI on the ESO 2.2m}
\end{deluxetable}

\clearpage

\begin{deluxetable}{cccccc}
\tablecaption{Values used to convert Vega magnitudes to the AB system$^a$.\label{tab:abconversions}}
\tablewidth{0pt}
\tablehead{
\colhead{$U$} & \colhead{$B$} & \colhead{$V$} & \colhead{$R$} & \colhead{$I$} & \colhead{$K$}
}
\startdata
+0.883 & -0.098 & +0.011 & +0.191 & +0.426 & +1.866\\
\enddata
\tablenotetext{a}{Conversions listed in the sense $m_{\rm AB}=m_{\rm VEGA} + C$.}

\end{deluxetable}

\clearpage

\begin{table}
{\scriptsize 
\begin{center}
  \caption{5-$\sigma$ Limiting magnitudes and seeing estimates for the PDS. \label{tab:optical_depths}}
  \begin{tabular}{ccccccccccccc}
\tableline\tableline
    Pointing & \multicolumn{2}{c}{U} &\multicolumn{2}{c}{B} & \multicolumn{2}{c}{V} & \multicolumn{2}{c}{R} & \multicolumn{2}{c}{I} &\multicolumn{2}{c}{K}   \\
    & 5-$\sigma$$^a$ & Seeing$^b$ & 5-$\sigma$ & Seeing & 5-$\sigma$ & Seeing & 5-$\sigma$ & Seeing & 5-$\sigma$ & Seeing & 5-$\sigma$ & Seeing \\
    \tableline
    3  & 24.78 & 1.32 & 24.64 & 1.50 & 24.30 & 1.45 & 24.00   & 1.49    & 24.39 & 1.42 & \nodata & \nodata\\
    7  & 25.10   & 1.22 & 24.95 & 1.53 & 24.32 & 1.48 & 24.61   & 1.21    & 24.67 & 1.20 & 21.92   & 0.97   \\
    11 & 25.11     & 1.17 & 24.55 & 1.45 & 24.52 & 1.12 & \nodata & \nodata & 22.84 & 1.26 & \nodata & \nodata\\
    \tableline
  \end{tabular}
\tablenotetext{a}{All magnitudes are listed in the AB system and are \textsc{magauto} \sex\ magnitudes}
\tablenotetext{b}{Seeing estimates are given in arcseconds}
\end{center}
}
\end{table}

\clearpage

\begin{table}
{\scriptsize 
\begin{center}
  \caption{Completeness magnitudes for the various optical/near-IR pointings\label{tab:completeness}}
  \begin{tabular}{ccccccccccccc}
\tableline\tableline
    Pointing & \multicolumn{2}{c}{U} &\multicolumn{2}{c}{B} & \multicolumn{2}{c}{V} & \multicolumn{2}{c}{R} & \multicolumn{2}{c}{I} &\multicolumn{2}{c}{K}   \\
    & 80\% & 50\% & 80\% & 50\% & 80\% & 50\% & 80\% & 50\% & 80\% & 50\% & 80\% & 50\% \\
    \tableline
    3  & 24.40 & 24.98 & 24.14 & 24.59 & 23.99 & 24.51 & 23.86   & 24.28   & 23.90 & 24.43 & \nodata & \nodata\\
    7  & 24.72 & 25.32 & 24.41 & 24.83 & 23.92 & 24.47 & 24.32   & 24.92   & 24.13 & 24.77 & 21.47 & 22.02\\
    11 & 24.67 & 25.30 & 24.10 & 24.50 & 24.03 & 24.58 & \nodata & \nodata & 22.76 & 23.07 & \nodata & \nodata\\
    \tableline
  \end{tabular}
\end{center}
}
\end{table}

\clearpage

\begin{deluxetable}{lcc}
\tablecaption{The results of cross-correlating the $\chi^2$ and $U$-band catalogs.\label{tab:radiomatches}}
\tablewidth{0pt}
\tablehead{
\colhead{Dataset} & \colhead{Number of radio objects observed} & \colhead{Number of radio objects detected}
}
\startdata
Optical ($\chi^2$) & 839 & 659 (79\%)\\
$U$-CTIO & 1063          & 569 (54\%)\\
$U$-ESO  & 256           & 53  (21\%)\\
$K$-band & 111           & 91  (82\%)\\
Total$^a$ & 1331 & 778 (58\%)\\
\enddata
\tablenotetext{a}{Includes all independently observed radio sources in all imaging campaigns}
\end{deluxetable}

\clearpage

\begin{deluxetable}{lcccccccc}
\tabletypesize{\scriptsize}
\tablewidth{0pt}
\tablecaption{Photometric data for the radio/optical identifications\label{tab:catalog}}
\tablehead{
\colhead{ID} &
 \colhead{$U$} &
 \colhead{$B$} &
 \colhead{$V$} &
 \colhead{$R$} &
 \colhead{$I$} &
 \colhead{$K$} &
\colhead{$z_{\mathrm{spec}}$} &
\colhead{$z_{\mathrm{phot}}$}
}
\startdata
PDSJ010922.1-451402 & \nodata &  23.48 $\pm$ 0.11 &  23.38 $\pm$ 0.18 &  23.38 $\pm$ 0.15 &  23.62 $\pm$ 0.20 & \nodata & \nodata & \nodata \\
PDSJ010926.9-450627 & \nodata &  20.14 $\pm$ 0.02 &  19.99 $\pm$ 0.02 &  19.92 $\pm$ 0.02 &  19.94 $\pm$ 0.02 & \nodata & \nodata & \nodata \\
PDSJ010927.0-452121 & \nodata &  18.80 $\pm$ 0.02 &  17.98 $\pm$ 0.02 &  17.71 $\pm$ 0.02 &  17.55 $\pm$ 0.02 & \nodata &  0.172 & \nodata \\
PDSJ010932.3-450444 & \nodata &  19.46 $\pm$ 0.02 &  18.47 $\pm$ 0.02 &  18.14 $\pm$ 0.02 &  17.99 $\pm$ 0.02 & \nodata & \nodata & \nodata \\
PDSJ010934.9-451646 & \nodata &  24.40 $\pm$ 0.17 &  24.00 $\pm$ 0.16 &  23.99 $\pm$ 0.12 &  24.05 $\pm$ 0.14 & \nodata & \nodata & \nodata \\
PDSJ010935.4-451301 & \nodata &  24.78 $\pm$ 0.31 &  23.90 $\pm$ 0.25 &  23.64 $\pm$ 0.16 &  23.01 $\pm$ 0.11 & \nodata & \nodata & \nodata \\
PDSJ010935.4-451853 & $>$24.98 &  23.72 $\pm$ 0.20 &  23.15 $\pm$ 0.17 &  22.96 $\pm$ 0.11 &  23.08 $\pm$ 0.13 & \nodata & \nodata & \nodata \\
PDSJ010935.6-460620 &  23.70 $\pm$ 0.22 & \nodata & $>$24.67 & \nodata & \nodata & \nodata & \nodata & \nodata \\
\enddata
\end{deluxetable}

\end{document}